\documentclass{aa}

\usepackage{booktabs}
\usepackage{threeparttable}
\usepackage{graphicx}
\usepackage{txfonts}
\usepackage{bm}
\usepackage{ulem}
\usepackage{hyperref}
\hypersetup{
    colorlinks=true,
    citecolor=blue,
    linkcolor=red,
    filecolor=magenta,
    urlcolor=cyan,
} 
\usepackage{siunitx}
\usepackage{makecell}
\usepackage{stfloats}
\usepackage{bbding}

\begin{document} 

    \title{Evidence for stellar contamination in the transmission spectra of HAT-P-12b}

    \author{C. Jiang \inst{1, 2}\fnmsep\thanks{\email{czjiang@pmo.ac.cn}}
            \and
            G. Chen \inst{1, 3}\fnmsep\thanks{\email{guochen@pmo.ac.cn}}
            \and
            E. Pall\'e \inst{4, 5}
            \and 
            F. Murgas \inst{4, 5}
            \and 
            H. Parviainen \inst{4, 5}
            \and
            F. Yan \inst{6}
            \and
            Y. Ma \inst{1}
            }
    
    \institute{CAS Key Laboratory of Planetary Sciences, Purple Mountain Observatory, 
        Chinese Academy of Sciences, Nanjing 210023, PR China
        \and
        School of Astronomy and Space Science, 
        University of Science and Technology of China, 
        Hefei 230026, PR China
        \and
        CAS Center for Excellence in Comparative Planetology, Hefei 230026, PR China
        \and
        Instituto de Astrof\'isica de Canarias, V\'ia L\'actea s/n, 
        E-38205 La Laguna, Tenerife, Spain
        \and
        Departamento de Astrof\'isica, Universidad de La Laguna, E-38206 La Laguna, Spain
        \and
        Institut f\"{u}r Astrophysik, Georg-August-Universit\"{a}t, Friedrich-Hund-Platz 1, D-37077 G\"{o}ttingen, Germany
        }

    \date{Received ...; accepted ...}
 
    \abstract
    {Transmission spectroscopy characterizes the wavelength dependence of transit depth,
    revealing atmospheric absorption features in planetary terminator regions.
    In this context, different optical transmission spectra of HAT-P-12b 
    reported in previous studies exhibited discrepant atmospheric features 
    (e.g., Rayleigh scattering, alkali absorption).}
    {We aim to understand the atmosphere of HAT-P-12b 
    using two transit spectroscopic observations by the Gran Telescopio CANARIAS (GTC),
    and to search for evidence of stellar activity
    contaminating the transmission spectra,
    which might be the reason behind the discrepancies.}
    {We used Gaussian processes to account for systematic noise 
    in the transit light curves  and used nested sampling for Bayesian inferences.
    We performed joint atmospheric retrievals
    using the two transmission spectra obtained by GTC OSIRIS,
    as well as previously published results,
    coupled with stellar contamination corrections for different observations.}
    {The retrieved atmospheric model exhibits no alkali absorption signatures, 
    but shows tentative molecular absorption features 
    including $\rm H_2O$, $\rm CH_4$ and $\rm NH_3$.
    The joint retrieval of the combined additional public data analysis
    retrieves similar results, but with a higher metallicity.}
    {Based on Bayesian model comparison, 
    the discrepancies of the transmission spectra of HAT-P-12b can be explained 
    by the effect of different levels of unocculted stellar spots and faculae.
    In addition, we did not find strong evidence for 
    a cloudy or hazy atmosphere from the joint analysis,
    which is inconsistent with prior studies based on 
    the observations of Hubble Space Telescope.}
    
    
    \maketitle
    
\section{Introduction}
    
    According to the NASA Exoplanet
    Archive\footnote{https://exoplanetarchive.ipac.caltech.edu} (as of July 2021),
    more than 4,400 exoplanets have been discovered, 
    among which only $\sim$100 have been observed using transmission spectroscopy.    
    A transmission spectrum presents the wavelength dependency 
    of planetary transit depth and may indicate 
    the atmospheric absorption and scattering features at the planetary terminator
    (\citealt{Seager2000}; \citealt{Brown2001}).
    By computing the radiative transfer of the stellar light 
    passing through the planetary atmosphere, 
    we can retrieve a one-dimensional atmospheric model
    that constrains its physical properties and chemical compositions,
    including the atmospheric temperature, metallicity, carbon-to-oxygen ratio, 
    abundances of atomic and molecular species and presence of clouds or hazes
    \citep{Madhusudhan2009}.
    
    The exoplanet HAT-P-12b is an interesting target
    and there are many related studies on its transmission spectroscopy
    \citep[e.g.,][]{Line2013,Mallonn2015,Sing2016,Barstow2017, Tsiaras2018,Alexoudi2018,Fisher2018,Pinhas2019,Deibert2019,Wong2020,Yan2020}.
    According to \cite{confirm-HATP12b} and \cite{Mancini2018}, 
    it is a low-density warm sub-Saturn with a mass of $0.201 \pm 0.012~M_{\rm J}$, 
    a radius of $0.919 \pm 0.023~R_{\rm J}$ and an equilibrium temperature of $955 \pm 11$ K.
    It orbits a K4 dwarf with a period of $\sim$3.21 days at a distance of $\sim$0.038 AU.
    The host star HAT-P-12 has a mass of $0.691 \pm 0.023~M_\odot$, 
    a radius of $0.679 \pm 0.013~R_\odot$ and an effective temperature of $4665 \pm 50$ K.
    \cite{Line2013} measured the near-infrared (NIR) transmission spectrum of HAT-P-12b 
    using the Hubble Space Telescope Wide Field Camera 3 (HST WFC3),
    which were observed before the implementation of the spatial scan mode \citep{Deming2013}.
    They failed to detect the expected water absorption.
    \cite{Sing2016} combined optical spectroscopy from 
    Space Telescope Imaging Spectrograph (HST STIS), NIR spectroscopy from HST WFC3, 
    and broad-band photometry at 3.6 $\rm \mu m$ and 4.5 $\rm \mu m$ 
    from the Infrared Array Camera (IRAC) of Spitzer Space Telescope.
    Their results showed a Rayleigh scattering slope 
    caused by haze aerosols and a possible potassium absorption signal.
    \cite{Wong2020} reanalyzed the data used in \cite{Sing2016} 
    coupled with two more transits observed in the spatial scan mode of the HST WFC3
    and two secondary eclipses using broad-band photometry from Spitzer IRAC.
    They found a cloudy and hazy atmosphere without alkali features.
    In contrast to the space-based observations, 
    the ground-based observations reported by \cite{Mallonn2015} 
    and \cite{Yan2020} presented relatively flat and featureless 
    transmission spectra in the optical, favoring a cloudy atmosphere.    
    \cite{Alexoudi2018,Alexoudi2020} proposed that 
    such discrepancy in the offsets and slopes of transmission spectra
    can be attributed to the inaccuracy of orbital geometry parameters 
    (semi-major axis, orbital inclination, or impact parameters) 
    used in the spectroscopic light curve fitting.    
    In addition, \cite{Deibert2019} performed high-resolution transit spectroscopy on HAT-P-12b,
    and they revealed a $3.2 \sigma$ detection of sodium absorption for the first time.
    
    In this study, we analyze the optical transmission spectra of HAT-P-12b 
    observed by the Optical System for Imaging and low-Intermediate-Resolution 
    Integrated Spectroscopy \citep[OSIRIS;][]{OSIRIS2000} at the Gran Telescopio CANARIAS (GTC). 
    We aim to validate the absorption signals of sodium and potassium
    and the Rayleigh scattering slope in the optical band,
    and to compare our results with previous observations using other instruments.

    In the next section, we summarize the observation details and data reduction processes. 
    Then we introduce our method of transit light curve analysis
    and present the transmission spectra in Sect. \ref{sect_transit_light_curve_analysis}. 
    The atmospheric retrieval is illustrated in Sect. \ref{sect_atmospheric_retrieval}. 
    We discuss other possible interpretations 
    of the transmission spectra in Sect. \ref{sect_discussion},
    in comparison with prior studies.
    We draw our conclusions in Sect. \ref{sect_conclusions}.   

\section{Observations and data reduction}
    \label{sect_observations_data_reduction}

    \subsection{Transit spectroscopy observations}
    \label{sect_observations}

    Two transit events of HAT-P-12b were observed by GTC OSIRIS 
    in the long-slit spectroscopy mode on the nights of 
    20 Apr 2012 (hereafter OB12) and 17 Mar 2013 (hereafter OB13), respectively.
    During the observations, one reference star GSC2 N130301284 
    was simultaneously observed for differential spectrophotometry.
    The target and the reference stars have comparable {\it r}-magnitudes,
    which are 12.33 and 12.84 \citep{URAT1}, respectively.
    They were aligned through a $12''$-wide and $7.4'$-long slit 
    perpendicular to the dispersion direction with a distance of $2.44'$.
    The pixel scale was $0.254''$ after a $2\times2$ pixel binning.
    The detector of OSIRIS consists of a mosaic of two Marconi CCDs
    ($1024\times2048$ pixels after $2\times2$ pixel binning) 
    with a $9.4''$ gap between them.
    In OB12, the target star was placed on CCD1 and the reference star was on CCD2, 
    while in OB13, both stars were placed on CCD2. 
    An R1000R grism was used to acquire the stellar spectra 
    with a dispersion of 2.62 \si{\angstrom} per pixel, 
    covering a wavelength range of 5,100 -- 10,000 \si{\angstrom}.
    More observation details are listed in Table \ref{table_ob_info}. 

    \begin{table}[htbp]
        \centering
        \begin{threeparttable}[b]
        \caption{Details of two transit observations for HAT-P-12b. }
        \begin{tabular}{lcc}
        \toprule
            Parameter   & OB12      & OB13 \\
        \midrule
            Program ID
                & GTC17-12A \tnote{a} 
                & GTC52-13A \tnote{a} \\
            RA (J2000)      
                & \multicolumn{2}{c}{$\mathrm{13^h 57^m 33.467^s}$} \\
            DEC (J2000)     
                & \multicolumn{2}{c}{$+43^\circ 29' 36.602''$} \\
            Observing date  
                & 21 Apr 2012  
                & 17 Mar 2013 \\
            Start time (UT)
                & 00:13:13 
                & 22:50:34 \\
            End time (UT)
                & 03:31:07
                & 02:34:42 \\
            Exposure time (s)   
                & 45 
                & 40 \\
            Readout velocity (kHz) 
                & 500 
                & 200 \tnote{b} \\
            Readout time (s)    
                & 7.8 
                & 21.0 \\
            Readout noise ($\rm e^-$) 
                & $\sim$8.0 
                & $\sim$4.5 \\
            Frame number    
                & 209 
                & 213 \\
            Airmass         
                & 1.03 -- 1.17 
                & 1.05 -- 1.79 \\
            FWHM \tnote{c} 
                & $0.64''$ -- $1.02''$ 
                & $0.94''$ -- $1.70''$  \\
            Resolving power (\si{\angstrom}) 
                & 6.6 -- 10.5 
                & 9.7 -- 17.5  \\
        \midrule
        \end{tabular}
        
        \begin{tablenotes}
            \item[a] PI: E. Pall{\'e}.
            \item[b] Readout speed of 200 kHz is currently
            the standard mode for OSIRIS.
            \item[c] Full width at half maximum (FWHM) 
            of the point spread function (PSF) of stars in the spatial direction
            at the central wavelength of 7187 \si{\angstrom}.
        \end{tablenotes}
        
        \label{table_ob_info}
        \end{threeparttable}
    \end{table}
    
    We note that the dome shutter of GTC could not be fully opened 
    before November 2015\footnote{
    http://www.gtc.iac.es/news/posts.php\#post\_2015\_2}.
    Therefore partial dome vignetting was likely presented for frames 
    acquired with elevations above $\sim$72 degrees\footnote{
    http://www.gtc.iac.es/instruments/osiris/media/OSIRIS-USER-MANUAL\_v3\_1.pdf}.
    This problem could introduce significant but smooth systematics 
    in the OB12 light curves,
    as it occurred from the $8^{\rm th}$ to the $114^{\rm th}$ frames.
    However, this problem was negligible for the OB13 dataset 
    because the corresponding elevations were lower than 72 degrees during the observation.
    
    In addition, a subset of the OB12 spectral images were contaminated by a moving ghost 
    that could originate from an instrumental internal light reflection.
    This ghost showed up in a total of 83 frames during OB12
    and overlapped a few pixels of the target's spectrum for 30 frames.
    The intensity of the ghost was two orders of magnitude 
    smaller than the peak values of stellar flux.
    For white-light curves, it has negligible influence after a broadband integration.
    But for spectroscopic light curves, it would cause significant flux anomalies
    when the ghost overlapped the target's spectrum.
    Figure \ref{fig_ghost} in the Appendix presents its shape and trace on CCD1,
    along with the resulting flux anomalies in the spectroscopic light curves.
    A similar case was reported by \cite{GTC-HATP32b} 
    in their GTC spectroscopic observations of HAT-P-32b.
    
    \subsection{Data reduction}
    
\begin{figure}
        \centering
        \includegraphics[width=\linewidth]{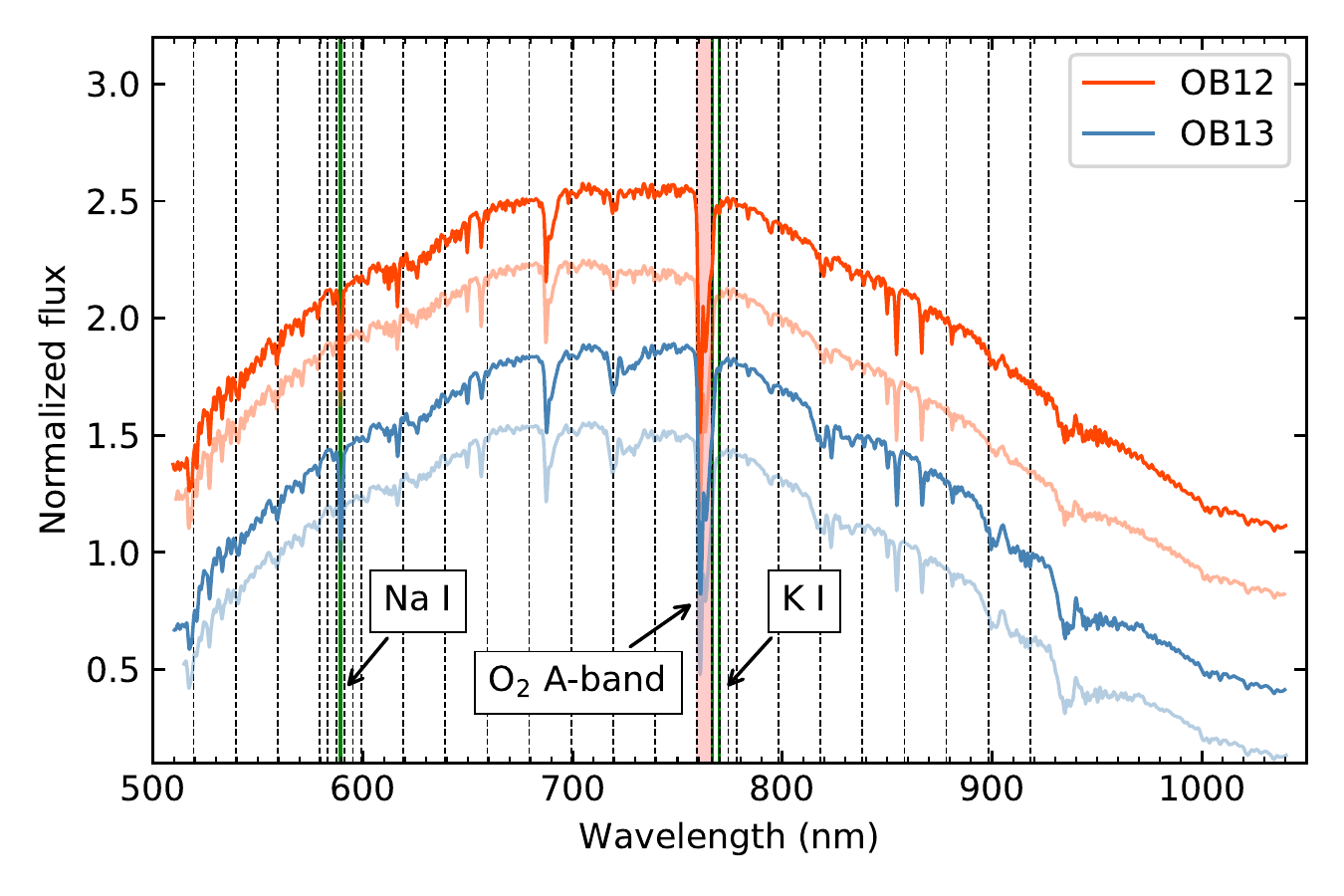}
        \caption{Stellar spectra of the target star HAT-P-12 (darker)
        and reference star GSC2 N130301284 (lighter) in OB12 (orange) and OB13 (blue). 
        The dashed lines are passbands for spectroscopic light curves.
        The green solid lines indicate the D-lines of Na and K.
        The red-shaded band is strongly affected by the telluric $\rm O_2$ absorption.
        The stellar fluxes have been normalized and arbitrarily shifted for clarity.}
        \label{fig_stellar_spectra}
    \end{figure}

    We reduced the spectral data following the procedures described in \cite{Chen2017a}.
    The frames of OB12 and OB13 were calibrated separately,
    but using the same procedures of \texttt{IRAF} programs \citep{code-IRAF},
    including corrections for overscan, bias, flat fields, cosmic rays and sky background. 
    The wavelength calibration was done utilizing the arc lines of HgAr, Xe and Ne
    observed through a narrower slit with the same R1000R grism.
    The best aperture size for extracting spectra was evaluated 
    by minimizing the scatter of white-light curves,
    and diameters of 25 pixels ($6.35''$) and 37 pixels ($9.40''$) 
    were chosen for OB12 and OB13, respectively.
    Figure \ref{fig_stellar_spectra} shows the stacked stellar spectra extracted 
    using the best aperture sizes.
    We adopted an overall 20-nm wavelength binning in the range 
    from 519.3 nm to 759.3 nm and from 766.2 nm to 918.2 nm.
    Narrower 4-nm bins were adopted centered around the sodium D-lines ($\sim$589.3 nm) 
    and the potassium D-lines ($\sim$768.2 nm).
    The light curve in the range of 759.3 -- 766.2 nm 
    had a large scatter and a lower signal-to-noise ratio
    due to the strong absorption of the telluric oxygen A-band (759 -- 770 nm),
    which is likely to impact the planetary transmission signal 
    at the potassium D2-line ($\sim$766.5 nm),
    while the telluric oxygen B-band ($\sim$689 nm) and water absorption ($\sim$820 nm)
    have negligible impacts after differential spectrophotometry.
    Therefore we excluded the narrow oxygen A-band in the light curve analyses.
    In addition, we also excluded the spectra with wavelength larger than 918.2 nm
    because of their low fluxes and strong fringing modulation at the red end.
    
\section{Transit light curve analysis} 
    \label{sect_transit_light_curve_analysis}

\subsection{White-light curves}

    The transit model was computed using \texttt{batman} \citep{code-batman}
    following the analytic model of \cite{Mandel2002}. 
    The free parameters of the transit model include 
    radius ratio $R_\mathrm p / R_\mathrm s$, 
    quadratic limb-darkening coefficients (LDCs) $u_1$ and $u_2$,
    orbital semi-major axis relative to stellar radius $a/R_\mathrm s$, 
    orbital inclination $i$, and central transit time $t_\mathrm c$.
    According to \cite{confirm-HATP12b},
    we adopted a circular orbit and used their transit ephemeris 
    to estimate the priors of $t_\mathrm c$:
    \begin{equation}
    \label{eq_ephemeris}
        t_\mathrm c = \mathrm{BJD_{TBD}~2454419.19556} + 3.2130598~n,   
    \end{equation}    
    where $n$ is the orbit number ($n=504$ for OB12; $n=607$ for OB13).
    
    We applied Gaussian process (GP) regression to predict 
    the systematic noise of transit light curves. 
    This method has been introduced in \cite{Gibson2012} and was further 
    applied in \cite{Pont2013}, \cite{Evans2016} and \cite{Sedaghati2017}, etc.
    We used \texttt{celerite} developed by \cite{code-celerite} 
    to implement fast one-dimensional GP regression.
    In our GP modeling, the normalized flux curve $\bm{f}(\bm t)$ 
    was fitted by the sum of a mean model $\bm{\mu} (\bm t; \bm\theta)$
    and a predictive curve $\bm g (\bm t; \bm\varphi)$ sampled from the GP model:
    \begin{equation}\label{eq_flux}
        \bm f(\bm t) = \bm\mu (\bm t; \bm\theta) 
        + \bm g(\bm t; \bm\varphi) + \bm\varepsilon,
    \end{equation}
    where $\bm\varepsilon$ is white noise, 
    $\bm t$ is time vector, 
    $\bm \theta$ and $\bm \varphi$ are free parameters of the mean model
    and kernel function, respectively. 
    The mean function $\bm\mu (\bm t)$ for white-light curves is just the transit model $m(t)$:
    \begin{equation}
        \mu(t; \bm{\theta}) = m(t; R_\mathrm p/R_\mathrm s, u_1, u_2, 
                                a/R_\mathrm s, i, t_\mathrm c),
    \end{equation}
    where $\bm{\theta} = ( R_\mathrm p/R_\mathrm s, u_1, u_2, a/R_\mathrm s, i, t_\mathrm c)$.
    The predictive curve $\bm g(\bm t)$ is sampled with a covariance matrix $\bm K$
    determined by a GP kernel function $k(\tau)$.
    \cite{Rasmussen2006} provides a detailed introduction to various GP kernels.
    Normally, a squared exponential (SE) kernel or a Mat\'ern class kernel 
    are flexible enough to account for the correlated noise.
    The former is infinitely differentiable and thus generates very smooth predictive curves,
    while the latter has variable levels of smoothness depending on its hyper-parameter $\nu$
    \citep[see Eq. 4.14 in][]{Rasmussen2006}.
    For large values of $\nu$, e.g., $\nu \ge 7/2$, 
    the corresponding GPs are indistinguishable from those generated by an SE kernel.
    In the ground-based transit observations, 
    the systematic noise may have rough changes
    due to seeing variation or pointing jitters.
    Therefore we selected the 3/2-order Mat\'ern kernel 
    ($\nu=3/2$) to fit the light-curve systematics, which has the form
    \begin{equation}
        k(\tau; \alpha, \ell) = \sigma_\mathrm{k}^2 \left( 1 + \frac{\sqrt{3} \tau}{\ell} \right) 
        \exp{ \left( - \frac{\sqrt{3} \tau}{\ell} \right)} ,
    \end{equation}
    where $\tau = |t_i - t_j|$ is the distance in time between two data points,
    $\sigma_\mathrm{k}^2$ and $\ell$ are the variance and length scale of the systematic noise.
    The estimated flux errors were added to the diagonal of the covariance matrix. 
    So the elements of the covariance matrix are
    \begin{equation}
        K_{ij} (k; \bm \varphi) = (\sigma_\mathrm{ph}^2+\sigma_\mathrm{add}^2)\delta_{ij}
        + k(|t_i - t_j|; \alpha, \ell),
    \end{equation}
    where $\bm \varphi = (\sigma_\mathrm{add}, \sigma_\mathrm{k}, \ell)$, 
    and $\delta_{ij}$ is the Kronecker delta function. 
    The error $\sigma_\mathrm{ph}$ is the photon-dominated noise 
    previously determined from reduced images,
    which underestimated the true white noise.
    Therefore, we used a free parameter $\sigma_\mathrm{add}$
    to account for additional jitters.

    We utilized nested sampling \citep{Skilling2004,Feroz2008}
    to estimate model evidence and parameter distributions in a Bayesian framework. 
    The package \texttt{PyMultiNest} \citep{code-PyMultiNest} was used, 
    which is a Python implementation of the code \texttt{MULTINEST} \citep{Feroz2009}. 
    In our light curve analyses, we calculated the parameter posteriors
    using 1,000 live points and a sampling efficiency of 0.3 in \texttt{PyMultiNest}.
    The nested sampling is terminated when the contributions of the remaining prior space 
    to $\ln\mathcal{Z}$ is less than 0.1.
    The posterior distributions are determined with $\sim$10,000 accepted samples 
    from a total of $\sim$600,000 samples.
    The parameter estimates are consistent in multiple runs.
    
    We adopted uniform priors on radius ratio $R_\mathrm p/R_\mathrm s$, 
    relative semi-major axis $a/R_\mathrm s$, orbital inclination $i$,
    and central transit time $t_\mathrm c$.
    The quadratic LDCs are difficult to constrain from the light curves. 
    Therefore, we assumed Gaussian priors for $u_1$ and $u_2$ based on the ATLAS model
    calculated by the code of \cite{code-limb-darkening} 
    using the stellar parameters of HAT-P-12.
    The GP parameters $\sigma_\mathrm{add}$, $\sigma_\mathrm{k}$, and $\ell$ 
    are constrained by log-uniform priors.
    
\begin{figure}
        \centering
        \includegraphics[width=\linewidth]{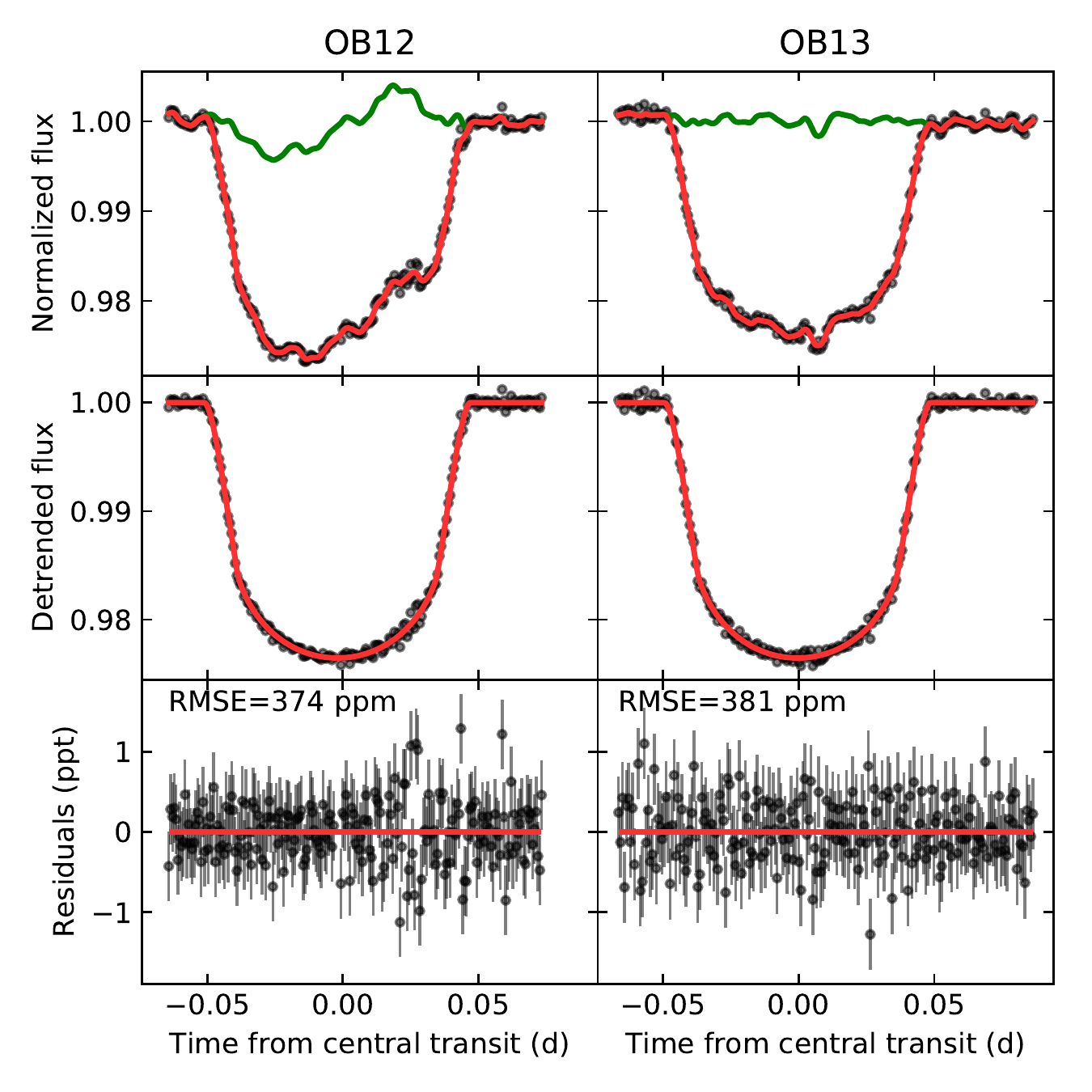}
        \caption{Joint white-light curve fitting of OB12 (left column) and OB13 (right column). 
        Top panels: observed white-light curves (black dots), 
        along with the best-fit results  (red lines)
        and systematics (green lines).
        Mid panels: detrended white-light curves (black dots)
        and the best-fit transit models (red lines).
        Bottom panels: residuals of light-curve fitting.}
        \label{fig_whiteLC}
    \end{figure}

    We used a single set of transit parameters (except $t_\mathrm{c}$)
    to fit the white-light transit model of OB12 and OB13,
    but separate GPs to fit the different systematics.
    The global log-likelihood function was the sum of 
    the log-likelihood for each GP.
    To demonstrate the consistency in transit parameters between two observations,
    we additionally performed individual analysis for each white-light curve.
    The white-light parameters derived from joint and individual analyses
    are listed in Table \ref{table_posterior}.
    The best-fit light curves and extracted systematics are shown in Fig. \ref{fig_whiteLC}.
    The results are consistent between OB12 and OB13,
    and also consistent with prior literature values of \cite{Wong2020}.
    
    \begin{table*}[htbp]
        \centering
        \renewcommand\arraystretch{1.2}
        \small
        \begin{threeparttable}[b]
        \caption{Transit parameters from broadband light curve fitting.}
        \begin{tabular}{llcccc}
        \toprule
            Parameters & Priors & Joint fit & OB12 & OB13 & \cite{Wong2020}\\
        \midrule
            Radius ratio, $R_\mathrm p/R_\mathrm s$ & 
                $\mathcal{U}(0.03, 0.3)$        & 
                $0.1387^{+0.0014}_{-0.0014}$    & 
                $0.1405^{+0.0036}_{-0.0035}$    &
                $0.1383^{+0.0016}_{-0.0016}$    &
                $0.13915^{+0.00053}_{-0.00054}$ \tnote{a}\\
            Relative semi-major axis, $a / R_\mathrm s$ & 
                $\mathcal{U}(2, 20)$ &
                $11.62^{+0.28}_{-0.28}$ &
                $11.84^{+0.48}_{-0.57}$ & 
                $11.70^{+0.25}_{-0.32}$ &
                $11.574^{+0.055}_{-0.054}$ \\
            Inclination, $i$ (deg)      & 
                $\mathcal{U}(80, 90)$   &
                $88.71^{+0.56}_{-0.39}$ &
                $88.99^{+0.72}_{-0.73}$ & 
                $88.86^{+0.65}_{-0.51}$ &
                $88.655^{+0.090}_{-0.084}$ \\
            Central transit time, $t_\mathrm c$ (day) \tnote{b} & 
                $\mathcal{U}(-0.05, 0.05)$ &
                \makecell[c]{$-0.00202^{+0.00050}_{-0.00051}$ (OB12) \\
                            $-0.00035^{+0.00021}_{-0.00021}$ (OB13)} &
                $-0.00233^{+0.00068}_{-0.00070}$    & 
                $-0.00034^{+0.00021}_{-0.00022}$    &
                $0.783203^{+0.000025}_{-0.000025}$  \\
            Quadratic LDCs, $u_1$ & 
                $\mathcal{N}(0.51, 0.1)$ \tnote{c} &
                $0.53^{+0.06}_{-0.06}$  &
                $0.50^{+0.08}_{-0.08}$  & 
                $0.53^{+0.06}_{-0.06}$  &
                -- \\
            Quadratic LDCs, $u_2$ & 
                $\mathcal{N}(0.17, 0.1)$ \tnote{c} &
                $0.15^{+0.09}_{-0.09}$  &
                $0.17^{+0.09}_{-0.09}$  & 
                $0.15^{+0.09}_{-0.08}$  &
                -- \\

        \midrule
        \end{tabular}
        \label{table_posterior}
        
        \begin{tablenotes}
            \item[a] Observed with the STIS G750L instrument.
            \item[b] Values of $t_\mathrm c$ for OB12 and OB13 were subtracted 
                by the predicted values given by Eq. \ref{eq_ephemeris}, 
                which are 2,456,038.577699 for OB12 
                and 2,456,369.522859 for OB13 in $\rm BJD_{TBD}$. 
                The value of \cite{Wong2020} was subtracted by 2,357,368.
            \item[c] Calculated from the ATLAS stellar spectral templates.
        \end{tablenotes}
        
        \end{threeparttable}
    \end{table*}
    
\subsection{Spectroscopic light curves}\label{sect_spectroscopic_light_curves}
    
    We fit for separate $R_\mathrm{p}/R_\mathrm{s}$, $u_1$ and $u_2$
    with each spectroscopic light curve.
    The wavelength-independent parameters ($a / R_\mathrm s$, $i$, $t_\mathrm c$)
    were fixed to the median estimates derived from 
    the joint analysis of white-light curves (Table \ref{table_posterior}).
    In addition, a baseline was added to the mean function 
    to account for the common-mode systematics.
    Some prior studies, e.g., \cite{Kreidberg2014Nature} and \cite{Gibson2017},
    would divide each spectroscopic light curve by the best-fit systematics model 
    to reduce the amplitudes of systematics and also avoid introducing additional parameters. 
    As discussed in \cite{Gibson2017}, 
    such a ``divide-white'' method is based on the assumption
    that systematics are mainly wavelength-independent.
    However, with regard to OB12 and OB13, 
    the spectroscopic systematics exhibited wavelength dependence 
    to some extent (Fig. \ref{fig_specLCs_OB12} and Fig. \ref{fig_specLCs_OB13}).
    In this case, the ``divide-white'' method would instead 
    introduce additional noise in those passbands with weaker systematics.
    Therefore, we allowed the common mode $w(t)$ 
    to have a variable amplitude $\alpha(\lambda)$,
    making it adaptive to the specific systematics in each passband.
    So the mean function for spectroscopic light curves became
    \begin{equation}\label{eq_mean_spectral}
        \mu (t; \bm\theta) = m(t; R_\mathrm p / R_\mathrm s, u_1, u_2) 
                            + \alpha w(t),
    \end{equation}
    where $\bm\theta=(R_\mathrm p / R_\mathrm s, u_1, u_2, \alpha)$.
    The common modes $w =f_\mathrm{white}-m_\mathrm{white}$ were determined 
    from the best-fit white-light curves for OB12 and OB13 separately.
    We assumed a uniform prior of $\mathcal{U}(0,3)$ for $\alpha$.
    The amplitude $\alpha$ is larger than one 
    when the spectroscopic systematics have larger amplitudes than $w(t)$
    and is close to zero when the common mode feature is quite weak.

    As mentioned in Sect. \ref{sect_observations}, 
    the observed spectra in OB12 were occasionally covered 
    by a moving ghost (Fig. \ref{fig_ghost}).
    The affected range is 520 -- 552 nm and 750 -- 825 nm in wavelength.
    It is difficult to fit these flux anomalies using GP or parametric baseline functions,
    for which we have to treat these points as outliers.
    We inspected all 30 frames where the ghost overlapped the stellar spectrum
    and discarded them in our spectroscopic light curve analysis.

    Figures \ref{fig_specLCs_OB12} and \ref{fig_specLCs_OB13} 
    show the fitting results of spectroscopic light curves of OB12 and OB13, respectively.
    It is evident that the spectroscopic systematics 
    have smaller amplitudes at the blue and red ends.
    We compared the Bayesian evidence between the ``divide-white'' method
    and the ``variable-common-mode'' method when fitting the spectroscopic light curves 
    and found stronger evidence for the latter in most wavebands (Fig. \ref{fig_lc_evidence}).
    Then we calculated the Pearson correlation coefficients 
    of the common-mode amplitudes $\alpha$ and the flux response in narrow passbands,
    which are 0.91 for OB12 and 0.85 for OB13,
    indicating that $\alpha$ is actually strongly 
    correlated with flux for both observations 
    (Fig. \ref{fig_variable_amplitudes}).
    Therefore, we consider that in the cases of OB12 and OB13, 
    the ``variable-common-mode'' method has a better performance than the ``divide-white'' method
    in reducing the spectroscopic systematics.

\begin{figure}[htbp]
        \centering
        \includegraphics[width=\linewidth]{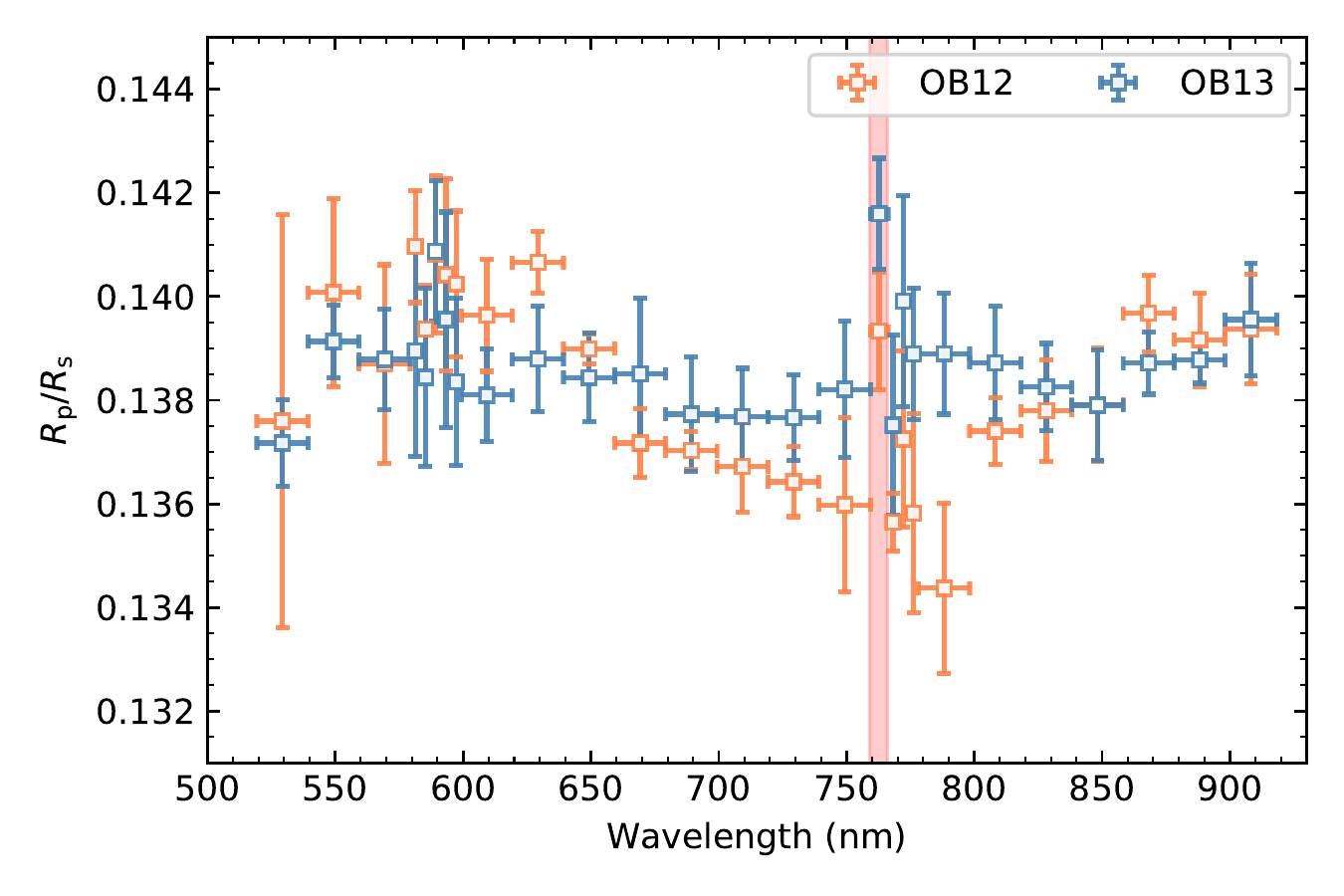}
        \caption{Transmission spectra of OB12 (orange) and OB13 (blue).
        The red-shaded passband indicates the telluric $\rm O_2$ A-band.}
        \label{fig_transpec}
    \end{figure}

\begin{figure}[htbp]
        \centering
        \includegraphics[width=\linewidth]{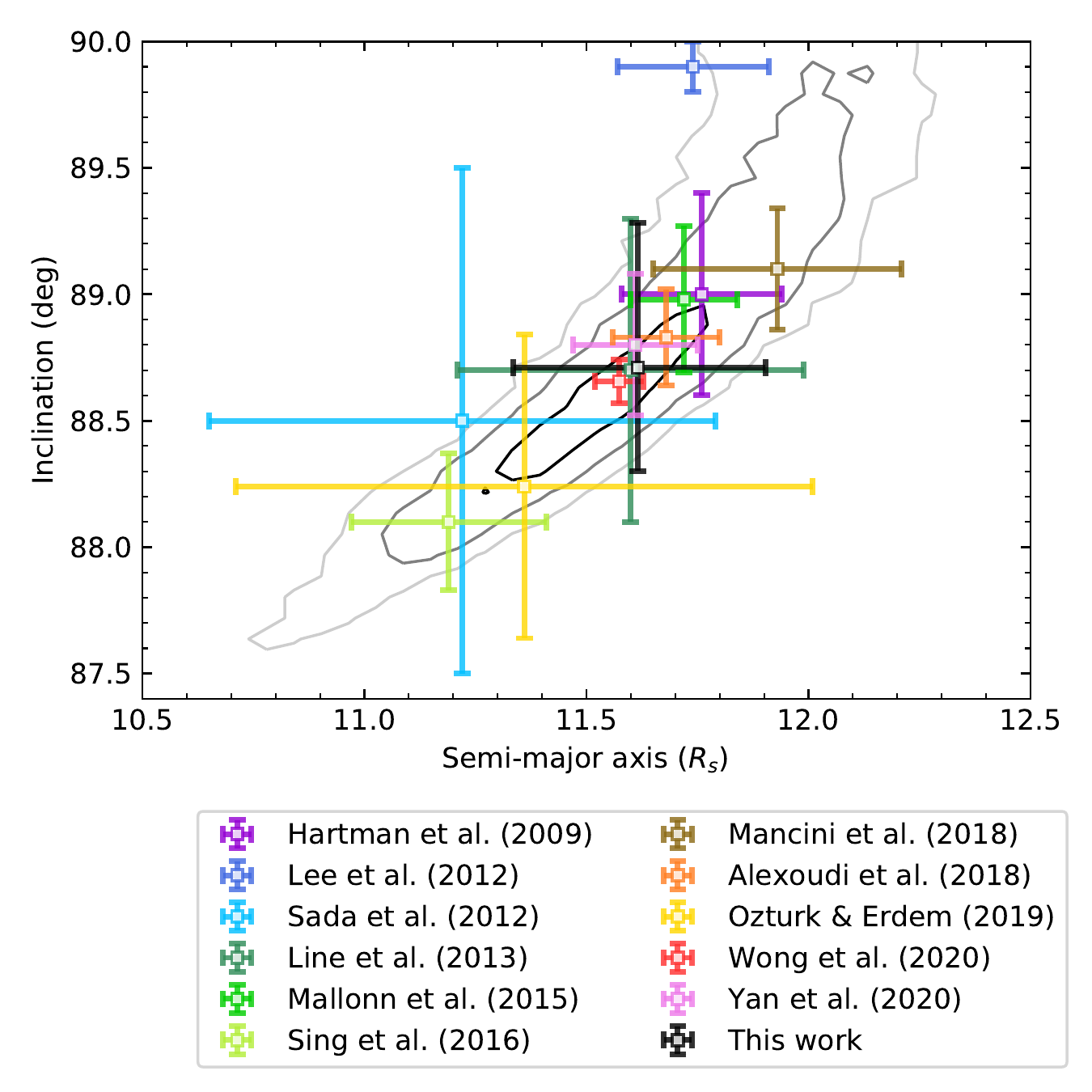}
        \caption{Posterior estimates of the orbital geometry parameters of HAT-P-12b
        compared with other literature values.
        The error bars indicate 1$\sigma$ credible intervals.
        The contours indicate 1 to 3$\sigma$ joint credible regions from nested sampling.
        }
        \label{fig_orbital_parameters}
    \end{figure}

    We present the transmission spectra of HAT-P-12b in Fig. \ref{fig_transpec}
    and list the posterior estimates of $R_\mathrm p / R_\mathrm s$ 
    in Table \ref{table_transit_depth}.
    The transmission spectrum of OB12 is slightly different from that of OB13
    and shows a larger slope in the range of 550 -- 800 nm.
    In Sect. \ref{sect_observations}, we mentioned that the OB12 data were affected by
    dome vignetting and instrumental internal light reflection.
    The effect of dome vignetting could introduce significant but smooth systematics
    to the light curves of both stars in OB12, which has been removed by GP regression
    and should not cause wavelength-dependent biases.
    Regarding the ``moving ghost'', only a small part of data points were affected
    and we have removed them as outliers (Fig. \ref{fig_ghost}),
    which should not be responsible for biases in a large wavelength range.
    \cite{Alexoudi2018} and \cite{Alexoudi2020} suggest 
    that a fixed but inaccurate orbital inclination or impact parameter
    will cause a different slope and a systematic offset in the optical transmission spectrum.
    As illustrated in Fig. \ref{fig_orbital_parameters},
    although there is a strong correlation 
    in the joint distributions of $a/R_\mathrm{s}$ and $i$, 
    our derived parameters are well consistent with other literature estimates.
    Furthermore, we applied the same values of $a/R_\mathrm{s}$ and $i$ 
    derived from the joint analysis of white-light curves 
    to all spectroscopic light curves of both observations.
    Thus the discrepancy between these two transmission spectra 
    should not be attributed to potential biases of orbital parameters.
    Instead we attribute such systematic bias to different levels 
    of stellar activity during OB12 and OB13,
    and interpret these two sets of results
    using the same planetary atmospheric model 
    but separate parameters accounting for stellar contamination.

    \begin{table}[htbp]
        \centering
        \renewcommand\arraystretch{1.2}
        \begin{threeparttable}[b]
        \caption{Radius ratios estimated from spectroscopic light curves.}
        \setlength{\tabcolsep}{4 mm}{
        \begin{tabular}{ccc}
        \toprule
            $\lambda$ (\si{\angstrom})          & 
            $R_\mathrm p/R_\mathrm s$ (OB12)    & 
            $R_\mathrm p/R_\mathrm s$ (OB13)    \\ 
        \midrule
        5193 -- 5393 & $0.1376 \pm 0.0040$ & $0.1372 \pm 0.0008$  \\
        5393 -- 5593 & $0.1401 \pm 0.0018$ & $0.1391 \pm 0.0007$  \\
        5593 -- 5793 & $0.1387 \pm 0.0019$ & $0.1388 \pm 0.0010$  \\
        5793 -- 5833 & $0.1410 \pm 0.0011$ & $0.1390 \pm 0.0020$  \\
        5833 -- 5873 & $0.1394 \pm 0.0008$ & $0.1384 \pm 0.0017$  \\
        5873 -- 5913 & $0.1408 \pm 0.0015$ & $0.1409 \pm 0.0014$  \\
        5913 -- 5953 & $0.1404 \pm 0.0019$ & $0.1396 \pm 0.0021$  \\
        5953 -- 5993 & $0.1402 \pm 0.0014$ & $0.1384 \pm 0.0016$  \\
        5993 -- 6193 & $0.1396 \pm 0.0011$ & $0.1381 \pm 0.0009$  \\
        6193 -- 6393 & $0.1407 \pm 0.0006$ & $0.1388 \pm 0.0010$  \\
        6393 -- 6593 & $0.1390 \pm 0.0003$ & $0.1384 \pm 0.0009$  \\
        6593 -- 6793 & $0.1372 \pm 0.0007$ & $0.1385 \pm 0.0015$  \\
        6793 -- 6993 & $0.1370 \pm 0.0004$ & $0.1377 \pm 0.0011$  \\
        6993 -- 7193 & $0.1367 \pm 0.0009$ & $0.1377 \pm 0.0009$  \\
        7193 -- 7393 & $0.1364 \pm 0.0007$ & $0.1377 \pm 0.0008$  \\
        7393 -- 7593 & $0.1360 \pm 0.0017$ & $0.1382 \pm 0.0013$  \\
        7593 -- 7662 & $0.1393 \pm 0.0011$ & $0.1416 \pm 0.0011$  \\
        7662 -- 7702 & $0.1356 \pm 0.0006$ & $0.1375 \pm 0.0017$  \\
        7702 -- 7742 & $0.1372 \pm 0.0017$ & $0.1399 \pm 0.0020$  \\
        7742 -- 7782 & $0.1358 \pm 0.0019$ & $0.1389 \pm 0.0013$  \\
        7782 -- 7982 & $0.1344 \pm 0.0016$ & $0.1389 \pm 0.0012$  \\
        7982 -- 8182 & $0.1374 \pm 0.0006$ & $0.1387 \pm 0.0011$  \\
        8182 -- 8382 & $0.1378 \pm 0.0010$ & $0.1383 \pm 0.0008$  \\
        8382 -- 8582 & $0.1379 \pm 0.0011$ & $0.1379 \pm 0.0011$  \\
        8582 -- 8782 & $0.1397 \pm 0.0007$ & $0.1387 \pm 0.0006$  \\
        8782 -- 8982 & $0.1392 \pm 0.0009$ & $0.1388 \pm 0.0004$  \\
        8982 -- 9182 & $0.1394 \pm 0.0011$ & $0.1396 \pm 0.0011$  \\
        \midrule
        \end{tabular}}
        \label{table_transit_depth}
        \end{threeparttable}
    \end{table}
    
\section{Atmospheric retrieval}
    \label{sect_atmospheric_retrieval}
    
    \subsection{Forward atmospheric modeling}

    We analyze the transmission signals of HAT-P-12b 
    using a one-dimensional isothermal model atmosphere 
    calculated by \texttt{PLATON} (\citealt{code-PLATON}; \citealt{code-PLATON2}).
    In \texttt{PLATON}, chromatic transit depths are mainly contributed 
    by gas absorption, collisional absorption, Mie scattering, 
    and the effect of unocculted stellar spots and faculae.    
    The atmospheric pressure is limited in range of $10^8$ -- $10^{-4}$ Pa
    and is equally divided into 500 layers in a logarithmic scale.
    The planetary radius at a reference pressure of $10^5$ Pa is a free parameter, 
    which determines the height of each atmosphere layer.
    The chemical abundances for calculating the absorption coefficients of each layer
    are based on the equilibrium chemistry model with 34 atomic and molecular species 
    computed by \texttt{GGChem} \citep{code-GGChem},
    which are: H, He, C, N, O, Na, K, 
    $\rm H_2$, $\rm H_2O$, $\rm CH_4$, $\rm CO$, 
    $\rm CO_2$, $\rm NH_3$, $\rm N_2$, $\rm O_2$, $\rm O_3$, $\rm NO$, 
    $\rm NO_2$, $\rm C_2H_2$, $\rm C_2H_4$,  
    $\rm H_2CO$, $\rm H_2S$, $\rm HCl$, $\rm HCN$, $\rm HF$, 
    $\rm MgH$, $\rm OCS$, $\rm OH$, $\rm PH_3$, 
    $\rm SiH$, $\rm SiO$, $\rm SO_2$, $\rm TiO$ and $\rm VO$.
    The sources of spectral line lists include
    ExoMol \citep{ExoMol}, HITRAN 2016 \citep{HITRAN2016},
    CDSD-4000 \citep{CDSD4000}, \cite{Rey2017}, and NIST \citep{NIST}.
    The collisional absorption coefficients are from HITRAN \citep{HITRAN2012}.
    The main parameters controlling chemical abundances are 
    atmospheric temperature ($T$), metallicity ($Z$) and carbon-to-oxygen ratio (C/O).
    The forward modeling of transmission spectra is based on the opacity line lists 
    with a resolution of $\lambda/\Delta\lambda=10,000$ as suggested by \cite{code-PLATON}.
    The high-resolution opacities were then integrated to the same wavelength bins 
    as those in Table \ref{table_transit_depth} 
    such that the likelihood function can be calculated.

    The modeling of atmospheric condensates considers Mie scattering.
    We assumed an optically thick cloud deck uniformly covering the planet
    and vertically extending to a pressure $P_\mathrm{cld}$, 
    below which the atmosphere is completely occulted by the opaque cloud
    while above atmospheric aerosols account for Mie scattering.
    As presented in the study of GJ 3470b by \cite{Benneke2019},
    Mie scattering can be used to characterize 
    scattering-induced opacities for all wavelengths,
    and it asymptotically approaches Rayleigh scattering 
    for particle sizes much smaller than the wavelength.
    \texttt{PLATON} uses the same algorithm as \texttt{LX-MIE} \citep{Kitzmann2018}
    to calculate the cross-sections of aerosol particles
    and allows the computation of Mie scattering 
    for three condensates ($\rm MgSiO_3$, $\rm SiO_2$, and $\rm TiO_2$) 
    using their actual wavelength-dependent refractive indices from \cite{Kitzmann2018}.
    We adopted $\rm MgSiO_3$, whose condensation temperature varies 
    from $\sim$1000 K at 1 Pa to $\sim$1700 K at $10^5$ Pa,
    as the major condensate for the atmosphere of HAT-P-12b, 
    although the results are statistically undistinguishable
    if the other two species were selected.
    The aerosol particle size is assumed to have a mean size $r_\mathrm{m}$ 
    and a standard deviation of 0.5 in the log-normal distribution.
    The vertical variation of aerosol number density is determined by the function
    $n_0\cdot \exp{(-h/H_\mathrm{aero})}$, 
    where $n_0$ is the maximum number density, 
    $h$ is the height above the cloud top,
    and $H_\mathrm{aero}$ is the aerosol scale height.
    Therefore, the free parameters for Mie scattering are
    the cloud-top pressure ($P_\mathrm{cld}$),
    the aerosol scale height relative to gas scale height 
    ($H_\mathrm{aero} / H_\mathrm{gas}$),
    the mean radius ($r_\mathrm m$) 
    and the maximum number density ($n_0$) of aerosol particles.

    \subsection{Stellar contamination correction}
    \label{sect_stellar_contamination}
    
    The photometric monitoring of HAT-P-12's activity 
    presented by \cite{Wong2020} covered the transit epochs of OB12 and OB13.
    The differential magnitudes in Cousins {\it R} band 
    were $-0.2689 \pm 0.0046$ and $-0.2708 \pm 0.0042$, 
    respectively in two observation seasons 
    (Sep. 2011  -- Jun. 2012; Sep. 2012  -- Jun. 2013), 
    but we cannot rule out the possibility of stellar activity 
    with {\it R}-band variation amplitudes less than the uncertainties.
    \cite{Johnson2021} performed forward modeling 
    on rotational variability in the Kepler and TESS bands,
    and found that the range of variability for K0 dwarfs
    can be less than 6 part per thousand when assuming a spot area coverage of 20\%
    and a temperature contrast of 560 K between spots and photosphere,
    while the presence of faculae has relatively little influence on the variability.
    Meanwhile, the light curve of OB13 exhibits dip-like features near the central transit, 
    which might be a hint of faculae occulted by the planet. 
    Therefore, we attempt to explain the discrepancy in transit depths between OB12 and OB13 
    with stellar contamination composed of spots and faculae.
    
    When a planet is transiting its host star, 
    the presence of unocculted spots and faculae 
    may cause significant wavelength-dependent offset of transit depth 
    over a large wavelength range (\citealt{McCullough2014}; \citealt{Rackham2018}; \citealt{Rackham2019}). 
    This is mainly contributed by the change of stellar spectrum
    due to the temperature contrast between
    active and quiescent areas on the photosphere.
    The observed transmission spectra will present positive offsets in spot-dominated circumstances
    but negative offsets in facula-dominated circumstances.
    In addition, these offsets are more significant at bluer wavelengths than at redder ends.
    Following \cite{code-PLATON}, we correct this effect by
    \begin{align} 
        \label{eq_stellar_contamination_1}
        & D'_\lambda = \frac{D_\lambda}{\beta_\lambda}, \\
        \label{eq_stellar_contamination_2}
        & \beta_\lambda = 1 + \phi_\mathrm{spot} 
            \left( \frac{S_\lambda(T_\mathrm{spot})} {S_\lambda(T_\mathrm{phot})}-1\right) 
            + \phi_\mathrm{facu} 
            \left( \frac{S_\lambda(T_\mathrm{facu})} {S_\lambda(T_\mathrm{phot})}-1\right),
    \end{align}
    where $D_\lambda = R_\mathrm{p}^2 / R_\mathrm{s}^2$ 
    is the transit depth calculated from the forward atmospheric model
    without considering stellar contamination, 
    $\beta_\lambda$ is the correction factor at wavelength $\lambda$,
    $S_\lambda(T)$ is the stellar spectra
    interpolated in the BT-NextGen (AGSS2009) stellar spectral grid \citep{Allard2012},
    $T_\mathrm{phot}$, $T_\mathrm{spot}$ and $T_\mathrm{facu}$ are 
    temperatures of the quiescent photosphere, spots and faculae, respectively,
    $\phi_\mathrm{spot}$ and $\phi_\mathrm{facu}$ are 
    the coverage fractions of spots/faculae.
    Other contributions such as limb-darkening of spots/faculae 
    or magnetic-dependent effects are not considered in this correction.
    
    We adopted the effective temperature of HAT-P-12 measured by \cite{Mancini2018}
    as an approximation of $T_\mathrm{phot}$ and fixed it to 4665 K.
    According to Fig. 7 in \cite{Berdyugina2005}, 
    the quiet-star-to-spot temperature contrast increases
    with stellar effective temperature 
    from about 200 K in M4V to 2000 K in G0V.
    Considering that HAT-P-12 is a K4 dwarf,  
    we assumed a uniform prior for $T_\mathrm{spot} - T_\mathrm{phot}$ 
    of $\mathcal{U}(-2000, -200)$.
    The temperature contrast between faculae and the photosphere 
    $T_\mathrm{facu} - T_\mathrm{phot}$
    was assumed to follow $\mathcal{U}(0, 1000)$.
    The coverage fractions $\phi_\mathrm{s}$ and $\phi_\mathrm{f}$ 
    are allowed to vary independently in a uniform prior of $\mathcal{U}(0, 0.5)$.
    The data of OB12 and OB13 were fitted by the same atmospheric model
    but were corrected by separate stellar contamination parameters
    ($T_\mathrm{spot}$, $T_\mathrm{facu}$, $\phi_\mathrm{spot}$, $\phi_\mathrm{facu}$).
    Thus, a total of eight free parameters were added to the retrieval algorithm.

    \subsection{Retrieval results}
    \label{sect_retrieval_results}
    
    The atmospheric retrieval was conducted with the nested sampling algorithm.
    We assumed uniform or log-uniform priors for all free parameters.
    The total log-likelihood is the sum of the log-likelihood function for each observation.
    We set 1,000 live points and a sampling efficiency of 0.3 in \texttt{PyMultiNest},
    and acquired $\sim$22,000 accepted samples 
    among a total of $\sim$1.3 million samples.
    The posterior estimates of all parameters are listed in Table \ref{table_retrieval_estimates}.
    The corresponding joint distributions are shown in Fig. \ref{fig_retrieval_corners}.
    The isothermal atmospheric temperature of $594^{+86}_{-59}$ K 
    is much lower than the equilibrium temperature of $955 \pm 11$ K.
    Similar low values were retrieved by \cite{Tsiaras2018} ($509 \pm 174$ K) 
    and \cite{Pinhas2019} ($456^{+70}_{-40}$ K),
    whereas other literature estimates are closer to the equilibrium temperature, 
    e.g., the effective temperature of $910^{+60}_{-70}$ K
    from the LBT observation by \cite{Yan2020}, 
    or the dayside blackbody temperature of $890^{+60}_{-70}$ K 
    from the secondary eclipse measurement of Spitzer IRAC by \cite{Wong2020}.
    \cite{MacDonald2020} proposed that a 1D retrieval model 
    applied on an inhomogeneous terminator atmosphere 
    could result in an underestimated atmospheric temperature.
    The atmospheric metallicity is estimated to be $\sim$10$^{0.63}$ times the solar value,
    while the metallicity of HAT-P-12 measured by \cite{Mancini2018} 
    is $[\mathrm{Fe/H}] = -0.20 \pm 0.09$.
    Therefore our estimated atmospheric metallicity is 
    approximately ten times that of the host star.
    Considering the low mass of HAT-P-12b ($M_\mathrm{p} \sim 0.2 M_\mathrm{J}$),
    this agrees with the mass-metallicity relation presented in \cite{Thorngren2016},
    which indicates that lower mass planets have weaker ability to accrete hydrogen and helium
    during their formation hence resulting in higher metallicity.
    The opaque cloud deck is found to exist at lower altitudes than previous literature estimates,
    and the cloud-top pressure is higher than $\sim$0.6 bar in a 95\% credible interval.
    The other parameters for Mie scattering 
    ($r_\mathrm{m}$, $H_\mathrm{aero}/H_\mathrm{gas}$, $n_0$) 
    have large uncertainties in line with the absence of scattering features.

    \begin{table*}[htbp]
        \centering
        \renewcommand\arraystretch{1.3}
        \begin{threeparttable}[b]
        \caption{Parameter estimates of the atmospheric retrieval.}
        \label{table_retrieval_estimates}
        \begin{tabular}{llcc}
        \toprule
            Parameter & Description & Prior range & Posterior estimates \\
        \midrule
            $R_0$ 
                & planet radius at pressure of 1 bar ($R_\mathrm{J}$)
                & $\mathcal{U}(0.8, 1.0)$ 
                & $0.8560^{+0.0077}_{-0.0063}$\\
            $T$
                & atmospheric temperature (K)
                & $\mathcal{U}(400, 2000)$ 
                & $594^{+86}_{-59}$\\
            $\log_{10} Z$
                & metallicity relative to solar
                & $\mathcal{U}(-1, 3)$
                & $0.63^{+0.33}_{-0.33}$ \\
            C/O
                & carbon-to-oxygen ratio
                & $\mathcal{U}(0.2, 2)$ 
                & $1.49^{+0.32}_{-0.42}$ \\
            $\log_{10} P_\mathrm{cld}$
                & cloud-top pressure (Pa)
                & $\mathcal{U}(-1, 8)$
                & $6.40^{+0.98}_{-1.00}$ \\
            $\log_{10} r_\mathrm{m}$ 
                & mean radius of aerosol particles ($\rm \mu m$)
                & $\mathcal{U}(-3, 2)$
                & $-0.78^{+1.67}_{-1.38}$ \\
            $\log_{10} H_\mathrm{aero}/H_\mathrm{gas}$
                & aerosol scale height relative to gas scale height
                & $\mathcal{U}(-3, 0)$
                & $-1.77^{+0.76}_{-0.74}$ \\
            $\log_{10} n_0$
                & aerosol number density at the cloud top ($\rm m^{-3}$)
                & $\mathcal{U}(0, 25)$
                & $14.04^{+6.46}_{-5.52}$ \\
            $\Delta T_\mathrm{spot}$
                & $T_\mathrm{spot} - T_\mathrm{phot}$ (K)
                & $\mathcal{U}(-2000, -200)$
                & $-308^{+64}_{-99}$ (OB12) | $-1134^{+279}_{-239}$ (OB13)\\
            $\Delta T_\mathrm{facu}$
                & $T_\mathrm{facu} - T_\mathrm{phot}$ (K)
                & $\mathcal{U}(0, 1000)$
                & $113^{+145}_{-74}$ (OB12) | $154^{+180}_{-93}$ (OB13) \\
            $\phi_\mathrm{spot}$
                & stellar spot coverage fraction (\%)
                & $\mathcal{U}(0, 50)$
                & $32.9^{+9.5}_{-8.9}$ (OB12) | $14.2^{+3.9}_{-2.8}$ (OB13)\\
            $\phi_\mathrm{facu}$
                & stellar facula coverage fraction (\%)
                & $\mathcal{U}(0, 50)$
                & $8.7^{+11.6}_{-6.2}$ (OB12) | $10.7^{+11.6}_{-6.5}$ (OB13)\\
        \midrule
        \end{tabular}
        
        \end{threeparttable}
    \end{table*}

\begin{figure}[htbp]
        \centering
        \includegraphics[width=\linewidth]{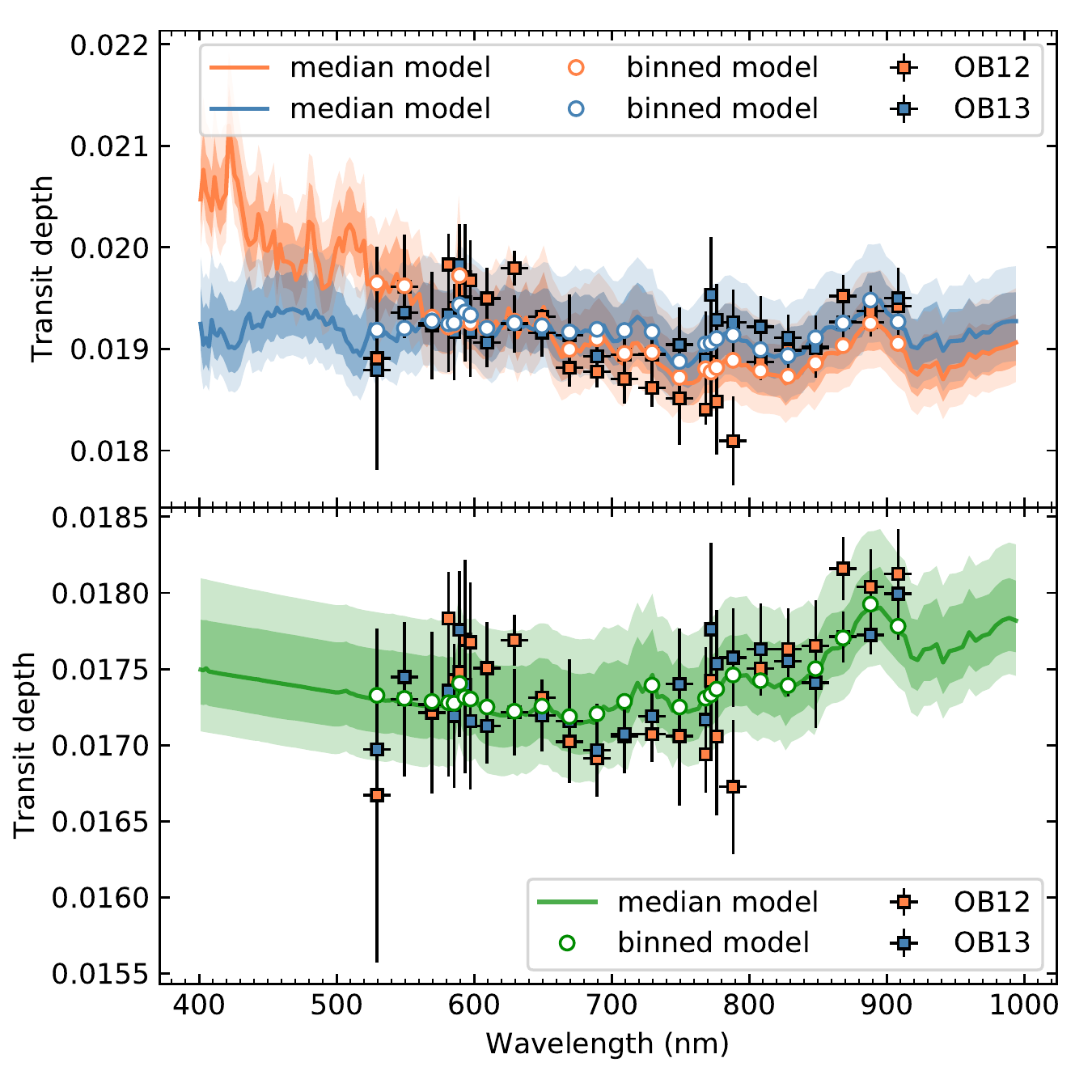}
        \caption{Retrieved transmission spectra of HAT-P-12b.
        Upper panel: retrieved models composed of 
        both model atmosphere and stellar contamination.
        Lower panel: transmission spectra and their retrieved models 
        after stellar contamination correction.
        The model transmission spectra are wavelength-binned to 
        a resolution of $\lambda/\Delta\lambda=200$.}
        \label{fig_retrieval_osiris}
    \end{figure}

    Figure \ref{fig_retrieval_osiris} illustrates the model transmission spectrum 
    derived from the joint retrieval after stellar contamination correction.    
    According to our stellar contamination model, 
    the transmission spectrum of OB12 exhibits 
    a lower spot-to-photosphere temperature contrast
    but a higher spot coverage fraction compared with those of OB13,
    which is due to the degeneracy between $T_\mathrm{spot}$ and $\phi_\mathrm{spot}$ 
    (Fig. \ref{fig_retrieval_corners}),
    whereas the retrieved facula temperatures and coverage fractions 
    are consistent between the two observations.
    The facula-to-spot area ratio ($Q=\phi_\mathrm{facu}/\phi_\mathrm{spot}$) are estimated to be 
    $0.27^{+0.36}_{-0.19}$ for OB12 and $0.73^{+0.77}_{-0.43}$ for OB13.
    According to \cite{Chapman1997}, the solar Q value was found to be 
    $16.7 \pm 0.5$ for a 7.5 year period during solar cycle 22 
    and increased as the solar cycle progressed.
    \cite{Herrero2016} suggests that this ratio 
    tends to be smaller for more active K and early-M dwarfs.
    Therefore, our results show that stellar activity during OB12 
    was slightly stronger than during OB13.
    However, the stellar activity index $\log R'_{\rm HK}$ of HAT-P-12b 
    was measured to be $-4.9\pm0.1$ by \cite{Mancini2018} using the HARPS-N spectra,
    which indicates low activity.
    We note that the spectra of HAT-P-12b in \cite{Mancini2018} 
    were observed in March and April 2015, more than two years after the OSIRIS observations.
    Thus it is possible that the level of stellar activity varied during this time.

\begin{figure}[htbp]
        \centering
        \includegraphics[width=\linewidth]{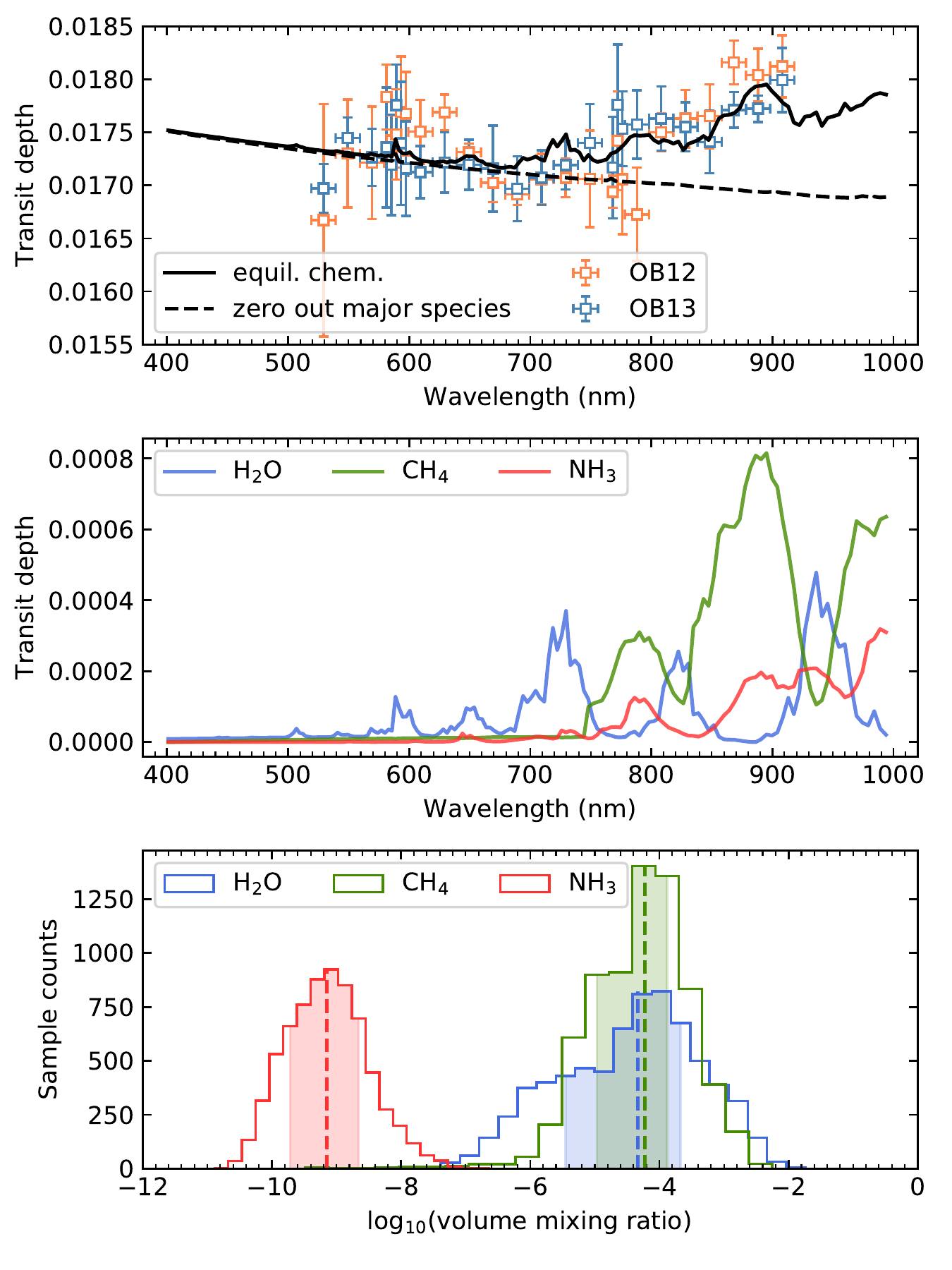}
        \caption{Major chemical species contributing to gas absorption features. 
        Top panel: the solid line is the median model 
        shown in the lower panel of Fig. \ref{fig_retrieval_osiris}.
        The dashed line is the same atmospheric model but assuming zero abundance 
        of $\rm H_2O$, $\rm CH_4$, and $\rm NH_3$.
        The transit depths of OB12 and OB13 have been corrected for stellar contamination.
        Middle panel: contributions of gas absorption to the chromatic transit depths.
        Bottom panel: posterior distributions of volume mixing ratios of major species.}
        \label{fig_compositions}
    \end{figure}

    The major species that contribute to gas absorption features
    are found to be $\rm H_2O$, $\rm CH_4$, and $\rm NH_3$ (Fig. \ref{fig_compositions})
    because of the high carbon-to-oxygen ratio of $1.5\pm0.3$,
    whose corresponding volume mixing ratios (in $\log_{10}$) are 
    $-4.3^{+1.0}_{-1.4}$, $-4.2^{+0.6}_{-0.9}$, and $-9.2^{+0.6}_{-0.6}$.
    Although other common species may also have 
    considerable mixing ratios under the assumption of equilibrium chemistry, 
    they have negligible absorption features in the optical
    thus are not presented in Fig. \ref{fig_compositions},
    including $\rm CO$ ($-3.1^{+0.5}_{-0.9}$), $\rm CO_2$ ($-6.1^{+1.0}_{-1.3}$), 
    $\rm N_2$ ($-3.6^{+0.3}_{-0.3}$), $\rm H_2S$ ($-4.0^{+0.4}_{-0.4}$), etc.
    The retrieved water abundance is consistent with prior literature values
    ($-3.61 \pm 1.48$, \citealt{Tsiaras2018}; 
    $-3.91^{+1.01}_{-1.89}$, \citealt{Pinhas2019}).    
    In addition, the chemistry model attributed the absorption peak at $\sim$590 nm 
    to the minor absorption from water rather than sodium.
    The alkali-metal atoms, including Na and K, are found to be 
    depleted in this atmosphere according to the equilibrium chemistry model, 
    whose abundances are approximately $3\times10^{-12}$ (Na) and $1\times10^{-13}$ (K).
    We note that \cite{Deibert2019} reported a possible detection 
    ($3.2\sigma$) of sodium absorption with high resolution transit spectroscopy, 
    which would indicate certain mechanisms 
    to keep the sodium atoms aloft in the upper atmosphere. 
    The high abundance of methane was never reported in prior studies,
    but it is favored by OSIRIS data to account for 
    the tentative absorption feature at $\sim890$ nm.

\section{Discussion}
    \label{sect_discussion}
    
    \subsection{Comparison with other atmospheric model assumptions}
    
    Here we examine other possible model assumptions 
    and compare all these models based on their Bayesian evidence.
    Following the notation of \cite{Feroz2009}, 
    the model evidence (or marginal likelihood) in Bayes' theorem is defined as 
    \begin{equation}
        \label{eq_evidence}
        \mathcal{Z} \equiv \mathrm{Pr}(\mathcal{D}|\mathcal{H}) = 
            \int \mathcal{L}(\bm{\varTheta}) \pi(\bm{\varTheta})
                \mathrm{d}^D \bm{\varTheta},
    \end{equation}
    where $\mathcal{D}$ is the data, $\mathcal{H}$ is the model hypothesis, 
    $\mathcal{L}$ is the likelihood function, $\pi$ is the prior function,
    $\bm{\varTheta}$ is the parameter vector and $D$ is its dimensionality.
    The evidence can be viewed as the average of the likelihood over the prior.
    With increasing dimensionality and broader prior space,
    the evidence will be exponentially weakened 
    unless the maximum likelihood rises considerably,
    which embodies Occam's razor.
    We then conducted model comparison between two hypotheses $\mathcal{H}_1$ and $\mathcal{H}_0$
    by calculating the Bayes factor:
    \begin{equation}
        \frac{\mathrm{Pr}(\mathcal{H}_1|\mathcal{D})}{\mathrm{Pr}(\mathcal{H}_0|\mathcal{D})}
        = \frac{\mathrm{Pr}(\mathcal{D}|\mathcal{H}_1)\mathrm{Pr}(\mathcal{H}_1)}
                {\mathrm{Pr}(\mathcal{D}|\mathcal{H}_0)\mathrm{Pr}(\mathcal{H}_0)}
        = \frac{\mathcal{Z}_1}{\mathcal{Z}_0},
    \end{equation}
    where the uninformative prior ratio 
    $\mathrm{Pr}(\mathcal{H}_1) / \mathrm{Pr}(\mathcal{H}_0)$
    for the two hypotheses is set to unity. 
    In practice, we calculate the log-evidence $\ln \mathcal{Z}$,
    and the logarithmic Bayes factor is simply 
    $\Delta \ln \mathcal{Z}_{10}=\ln \mathcal{Z}_1 - \ln \mathcal{Z}_0$.
    We interpret the Bayes factor based on the categories proposed by \cite{Kass1995}. 
    When $-1<\Delta \ln \mathcal{Z}_{10}<0$, there is weak evidence for $\mathcal{H}_0$ against $\mathcal{H}_1$.
    When $\Delta \ln \mathcal{Z}_{10}<-5$, 
    there is very strong evidence for $\mathcal{H}_0$ against $\mathcal{H}_1$.
    
    When considering other potential model combinations,
    we mainly focus on the contributions from gas absorption,
    condensate scattering, and stellar contamination.
    Therefore a total of eight hypotheses can be proposed and evaluated by nested sampling.
    The simplest model is a pure atmosphere 
    consisting of only hydrogen and helium ($\mathcal{H}_7$),
    where the $\rm H_2$-$\rm H_2$ and $\rm H_2$-He 
    collision-induced opacity and Rayleigh scattering are still considered,
    and it only has two free parameters: 
    the reference planetary radius $R_0$ and atmospheric temperature $T$.
    The most complicated model is the full model ($\mathcal{H}_0$) 
    proposed in Sect. \ref{sect_retrieval_results},
    which has 16 free parameters, eight for the atmospheric model 
    and the other eight for stellar contamination correction.
    Since the estimated Bayesian evidence for $\mathcal{H}_0$ 
    could reach a precision of 0.12 when using 1,000 live points 
    and a sampling efficiency of 0.3 in \texttt{PyMultiNest}, 
    we consider such hyperparameter settings are also suitable 
    for other simpler model hypotheses.
    The actual computational cost largely depends on the specific model,
    from $\sim$45 minutes for a null model $\mathcal{H}_7$ 
    up to $\sim$64 hours for a full model $\mathcal{H}_0$ with 32-core parallel computing 
    (3.00 GHz CPUs).
    
    Table \ref{table_model_comparison} lists all eight hypotheses
    and their estimated model evidence. 
    The model without condensate ($\mathcal{H}_2$)
    is calculated to have the largest model evidence.
    Comparing it with other models considering stellar contamination corrections
    ($\mathcal{H}_0$, $\mathcal{H}_1$ and $\mathcal{H}_3$),
    there is quite weak evidence for condensate scattering
    but very strong evidence for gas absorption. 
    For those models excluding stellar contamination
    ($\mathcal{H}_4$, $\mathcal{H}_5$, $\mathcal{H}_6$ and $\mathcal{H}_7$),
    their model evidences are considerably smaller than $\mathcal{H}_2$.
    Therefore, we can conclude that there is very strong evidence for the transmission spectra
    being affected by unocculted stellar spots and faculae
    but weak evidence for the presence of high-altitude clouds and hazes.
    This inference is different from other research papers on the atmosphere of HAT-P-12b. 
    Though the stellar contamination model favored by Bayesian model comparison
    can explain a sufficient degree of variability in the data,
    it is not the only explanation for the change of transmission spectra.
    There are also other possible effects that might cause such variability or inconsistency,
    e.g., instrumental systematics and telluric effects.

    \begin{table*}[htbp]
        \centering
        \renewcommand\arraystretch{1.2}
        \begin{threeparttable}[b]
        \caption{Bayesian evidence under different atmospheric model assumptions.}
        \label{table_model_comparison}
        \begin{tabular}{lcccccc}
        \toprule
            Hypothesis & 
            Gas absorption \tnote{a} & 
            Condensates \tnote{b} & 
            Stellar contamination \tnote{c} & 
            $D$ \tnote{d} & 
            $\ln \mathcal{Z}$ &
            $\Delta \ln \mathcal{Z}$ \tnote{e} \\
        \midrule
            $\mathcal{H}_0$: full model & 
                $\checkmark$ & $\checkmark$ & $\checkmark$ & 16 & $323.43 \pm 0.12$ & 0 \\
            $\mathcal{H}_1$: no gas absorption & 
                $\times$ & $\checkmark$ & $\checkmark$ & 14 & $309.35 \pm 0.12$ & -14.08 \\   
            $\mathcal{H}_2$: no condensates & 
                $\checkmark$ & $\times$ & $\checkmark$ & 12 & $324.00 \pm 0.12$ & 0.57 \\
            $\mathcal{H}_3$: only contamination & 
                $\times$ & $\times$ & $\checkmark$ & 10 & $307.56 \pm 0.11$ & -15.87 \\
            $\mathcal{H}_4$: no contamination & 
                $\checkmark$ & $\checkmark$ & $\times$ & 8 & $315.25 \pm 0.10$ & -8.18 \\
            $\mathcal{H}_5$: only condensate & 
                $\times$ & $\checkmark$ & $\times$ & 6 & $312.39 \pm 0.07$ & -11.04 \\
            $\mathcal{H}_6$: clear atmosphere &
                $\checkmark$ & $\times$ & $\times$ & 4 & $318.80 \pm 0.09$ & -4.63 \\
            $\mathcal{H}_7$: pure $\rm H_2$ and He & 
                $\times$ & $\times$ & $\times$ & 2 & $310.44 \pm 0.08$ & -12.99 \\
        \midrule
        \end{tabular}
        \begin{tablenotes}
            \item[a] Two additional parameters: $Z$ and C/O.
            \item[b] Four additional parameters: 
                $P_\mathrm{cld}$, $r_\mathrm{m}$, 
                $H_\mathrm{aero} / H_\mathrm{gas}$, and $n_0$.
            \item[c] Four additional parameters for each observation: 
                $\Delta T_\mathrm{spot}$, $\Delta T_\mathrm{facu}$, 
                $\phi_\mathrm{spot}$, and $\phi_\mathrm{facu}$.
            \item[d] Model dimensionality.
            \item[e] Logarithmic Bayes factor, 
                $\Delta \ln \mathcal{Z}_{k0}=\ln \mathcal{Z}_k - \ln \mathcal{Z}_0$.
        \end{tablenotes}
        \end{threeparttable}
    \end{table*}
    
\subsection{Comparison with the published data}

\begin{figure}[htbp]
        \centering
        \includegraphics[width=\linewidth]{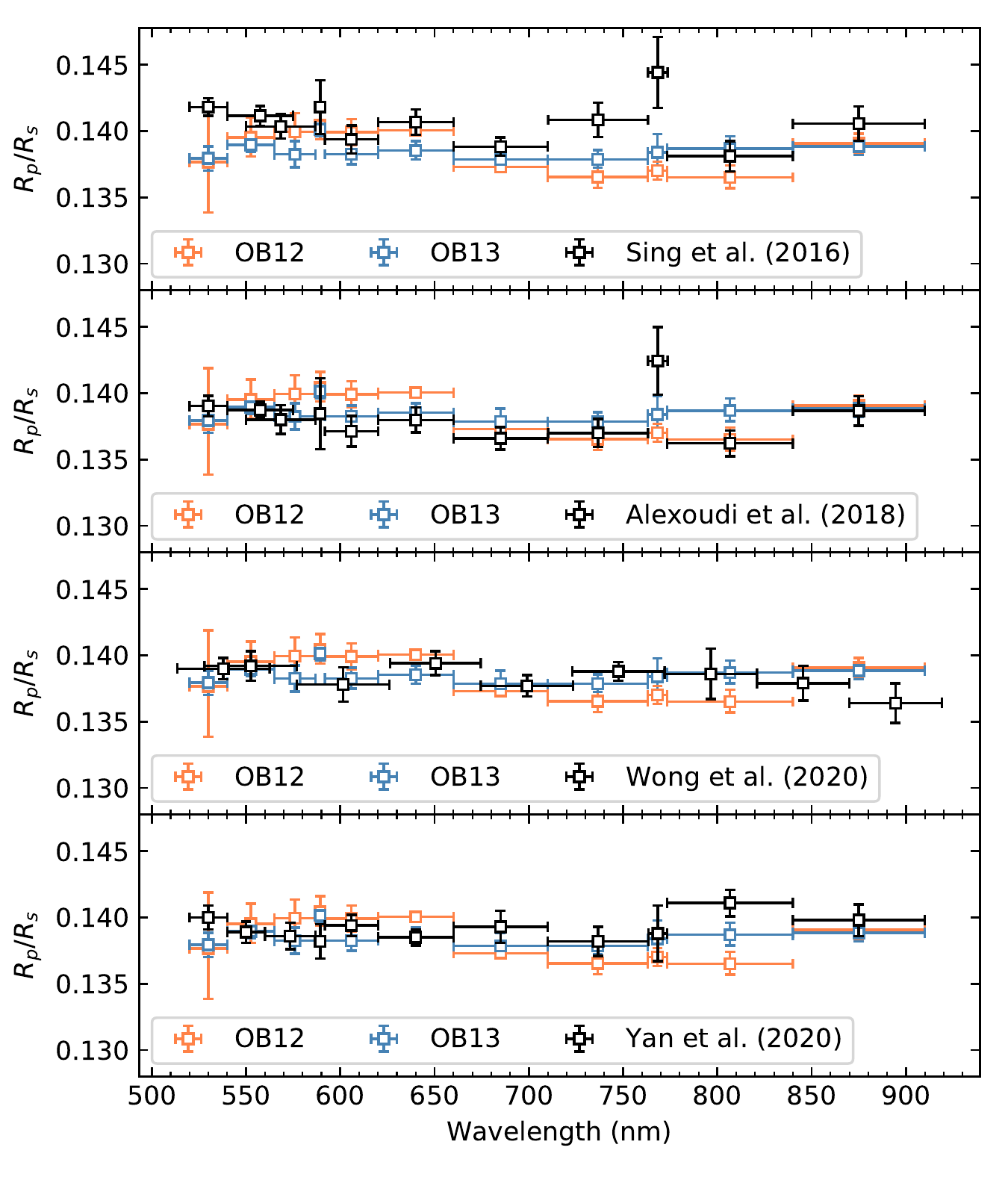}
        \caption{Transmission spectra of OB12 and OB13 recalculated 
        with similar wavelength bins used in other research papers.}
        \label{fig_compare_HST}
    \end{figure}

    To examine the consistency among the optical transmission spectra
    observed by other instruments, 
    we recalculate the transmission spectra of OB12 and OB13 
    using similar wavelength bins with those of 
    \cite{Sing2016} (S16), \cite{Alexoudi2018} (A18), 
    \cite{Wong2020} (W20) and \cite{Yan2020} (Y20),
    which are presented in Fig. \ref{fig_compare_HST}.
    The optical spectra of S16, A18 and W20 were derived from the same data set of HST STIS,
    including the observations of two transits performed by STIS G430L grating
    and one transit performed by STIS G750L grating.
    The spectrum of Y20 was observed by the dual-channel mode of 
    multi-object double spectrograph (MODS) on the Large Binocular Telescope (LBT),
    and Fig. \ref{fig_compare_HST} shows the weighted means of MODS1 and MODS2.
    Our results are close to those of A18, W20 and Y20,
    while the values of S16 basically lie above our two sets of spectra.
    The discrepant results of S16 and A18 on the mean and slope of the spectra
    are mainly due to the discrepant values of orbital geometry parameters 
    (Fig. \ref{fig_orbital_parameters}),
    but they both present a tentative detection of neutral potassium,
    which is however not detected or reported by other research, 
    including the high-resolution observation of \cite{Deibert2019}.

\subsection{Joint retrieval with the published data}

    We can combine the OSIRIS transmission spectra with 
    other transmission spectra to perform 
    a joint retrieval considering stellar contamination, 
    including data from \cite{Yan2020} (LBT MODS)
    and \cite{Wong2020} (HST STIS G750L, HST WFC3 G141 and Spitzer IRAC).
    We note that \cite{Wong2020} performed a joint analysis of transit light curves based on
    two transits observed by HST WFC3 in the scan mode (12 Dec 2015 and 31 Aug 2016) and 
    another two transits observed by Spitzer IRAC 
    (3.6 $\mu$m on 8 Mar 2013 and 4.5 $\mu$m on 11 March 2013).
    The stellar contamination correction of Eq. \ref{eq_stellar_contamination_2} 
    should not be applied on these multi-epoch NIR data sets.
    Fortunately, according to Fig. 3 of \cite{Pinhas2018},
    stellar contamination has weaker effects in NIR wavebands than in the optical
    and can be simply approximated as a transit depth offset or a rescaling factor.
    Furthermore, such a rescaling amplitude is within 1\% for the IRAC wavebands,
    for which we can neglect the effect of stellar contamination on the IRAC data.
    Regarding the data from WFC3 G141, we replaced $\beta_\lambda$ 
    in Eq. \ref{eq_stellar_contamination_1} with a free parameter $\beta_{\rm WFC3}$ 
    to apply independent rescaling of their transit depths.
    The transmission spectra of LBT MODS and HST STIS G750L
    were derived from single transit observations and contained sufficient data points
    (11 points for MODS and 10 points for G750L),
    and thus we can apply the same correction method as we did for the OSIRIS data.     
    In addition, although there were two observations 
    by HST STIS G430L presented in \cite{Wong2020},
    they were excluded in this joint retrieval for two reasons.
    One is that only four transit depths were available 
    for each observation of STIS G430L,
    which failed to constrain the atmospheric model 
    after a four-parameter stellar contamination correction
    ($\Delta T_\mathrm{spot}$, $\Delta T_\mathrm{facu}$, 
    $\phi_\mathrm{spot}$, and $\phi_\mathrm{facu}$).
    The other reason is that the rescaling approximation 
    is unsuitable for the NUV-to-optical bands
    where the stellar contamination effect is highly wavelength dependent,
    as it has been shown in Fig. \ref{fig_retrieval_osiris}, 
    as well as Fig. 3 of \cite{Pinhas2018}.

\begin{figure*}[htbp]
        \centering
        \includegraphics[width=\linewidth]{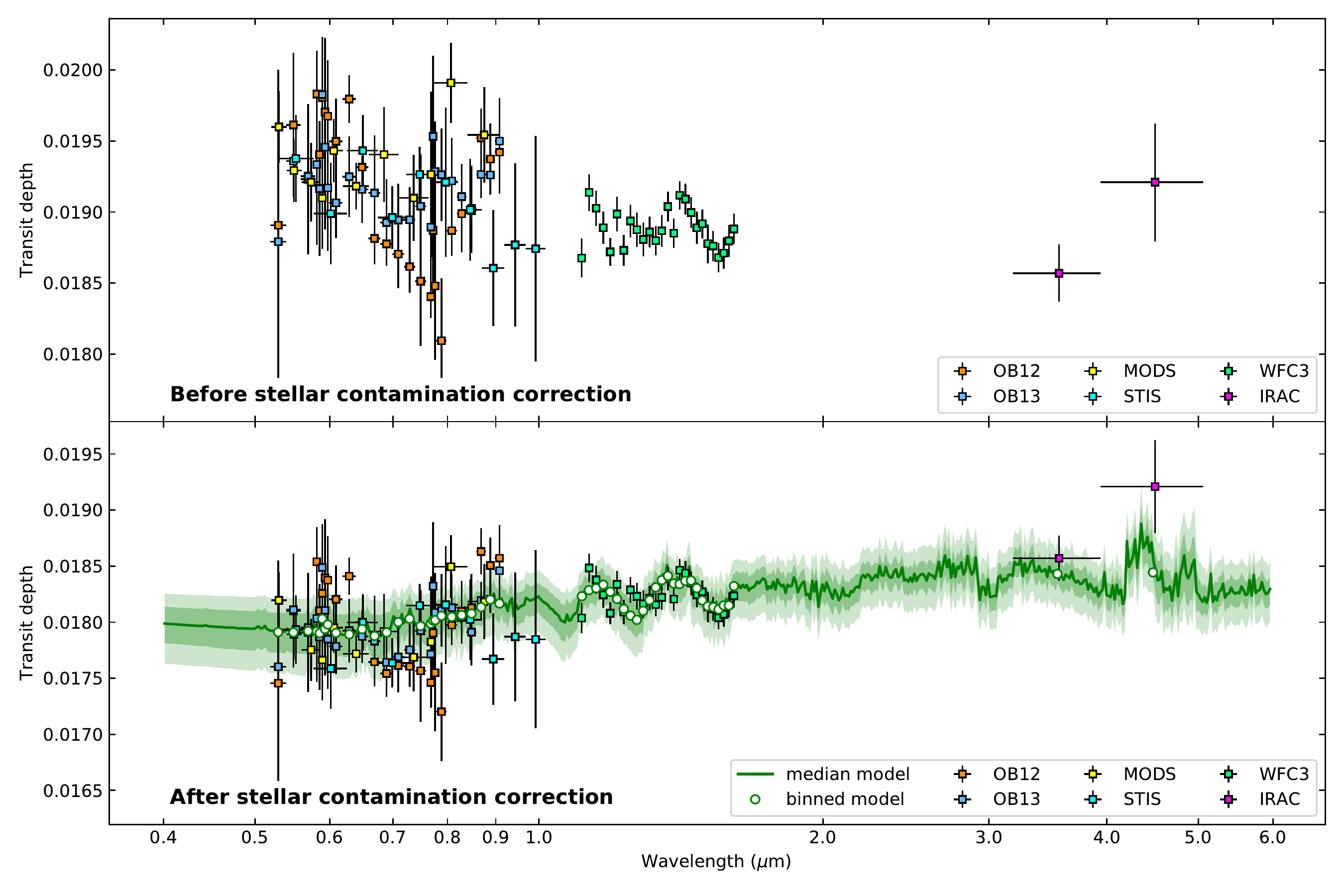}
        \caption{Retrieval analysis joint with the transit depths 
        presented in \cite{Yan2020} (LBT MODS) 
        and \cite{Wong2020} (HST STIS G750; HST WFC3 G141; Spitzer IRAC).
        The upper panel shows the raw transmission spectra of the six data sets.
        The lower panel shows the joint fit results with 
        the effects of stellar contamination corrected for each data set.
        The green solid line is the median model with 
        a resolution of $\lambda/\Delta\lambda=200$.
        The shaded areas indicate 68\% and 95\% credible intervals.
        The white dots are best-fit values calculated 
        at the same wavelength bins as the corresponding data points.
        In the range of 0.5 -- 1.0 $\mu$m, only the best-fit points 
        corresponding to OSIRIS wavebands are displayed for clarity.}
        \label{fig_retrieval_extra}
    \end{figure*}

    Figure \ref{fig_retrieval_extra} shows the results of this joint retrieval,
    where the posterior estimates of corresponding model parameters are listed 
    in Table \ref{table_retrieval_estimates_extra}.
    To reduce the computational cost, we adopted the pre-computed line lists 
    with a lower resolution of $\lambda/\Delta\lambda=1000$ in this part.
    The model inference is basically consistent with our previous analysis
    using OSIRIS data alone.
    There are evident gas absorption features in the optical and NIR wavebands.
    Although the signatures of methane are disfavored by the IRAC data,
    the atmospheric model still exhibits a low temperature of $570^{+86}_{-78}$ K
    and a high C/O ratio of $1.26^{+0.44}_{-0.49}$.
    The metallicity is found to be $\sim10^2$ times solar,
    which is higher than our previous estimate but consistent with that of \cite{Wong2020}.
    The model of condensate scattering is still poorly constrained,
    indicating low evidence for high-altitude clouds and hazes. 
    We also showed that the discrepancy between different transmission spectra 
    can be reduced after the stellar contamination corrections.
    A direct joint atmospheric retrieval without considering this effect 
    could be less reliable when the data in different wavebands 
    were acquired at different transit epochs.
    Therefore, it is crucial to perform transit spectroscopy
    that covers a broader wavelength range or even multiple bands in just one transit,
    which may be achieved by future telescopes such as James Webb Space Telescope 
    (JWST; e.g., \citealt{Greene2016}, \citealt{Schlawin2018}).
    Meanwhile, it is also necessary to repeat the spectroscopic observations
    at different transit epochs to check whether or not 
    the transmission spectra are contaminated by stellar activity.
    
    \begin{table*}[htbp]
        \centering
        \renewcommand\arraystretch{1.2}
        \begin{threeparttable}[b]
        \caption{Parameter estimates of atmospheric retrieval joint with other transmission spectra.}
        \label{table_retrieval_estimates_extra}
        \begin{tabular}{llcc}
        \toprule
            Parameter & Description & Prior range & Posterior estimates \\
        \midrule
            $R_0$ 
                & planet radius at pressure of 1 bar ($R_\mathrm{J}$)
                & $\mathcal{U}(0.8, 1.0)$ 
                & $0.8790^{+0.0049}_{-0.0060}$\\
            $T$
                & atmospheric temperature (K)
                & $\mathcal{U}(400, 2000)$ 
                & $570^{+86}_{-78}$\\
            $\log_{10} Z$
                & metallicity relative to solar
                & $\mathcal{U}(-1, 3)$
                & $2.23^{+0.17}_{-0.17}$ \\
            C/O
                & carbon-to-oxygen ratio
                & $\mathcal{U}(0.2, 2)$ 
                & $1.26^{+0.44}_{-0.49}$ \\
            $\log_{10} P_\mathrm{cld}$
                & cloud-top pressure (Pa)
                & $\mathcal{U}(-1, 8)$
                & $5.82^{+1.4}_{-1.4}$ \\
            $\log_{10} r_\mathrm{m}$ 
                & mean radius of aerosol particles ($\rm \mu m$)
                & $\mathcal{U}(-3, 2)$
                & $-0.61^{+1.55}_{-1.50}$ \\
            $\log_{10} H_\mathrm{aero}/H_\mathrm{gas}$
                & aerosol scale height relative to gas scale height
                & $\mathcal{U}(-3, 0)$
                & $-1.70^{+0.81}_{-0.81}$ \\
            $\log_{10} n_0$
                & aerosol number density at the cloud top ($\rm m^{-3}$)
                & $\mathcal{U}(0, 25)$
                & $11.83^{+7.54}_{-6.71}$ \\
            \\ \multicolumn{4}{c}{GTC OSIRIS (21 Apr 2012)} \\
            $\Delta T_\mathrm{spot}$ 
                & $T_\mathrm{spot} - T_\mathrm{phot}$ (K)
                & $\mathcal{U}(-2000, -200)$
                & $-300^{+68}_{-91}$\\
            $\Delta T_\mathrm{facu}$
                & $T_\mathrm{facu} - T_\mathrm{phot}$ (K)
                & $\mathcal{U}(0, 1000)$
                & $111^{+132}_{-71}$  \\
            $\phi_\mathrm{spot}$ 
                & stellar spot coverage fraction (\%)
                & $\mathcal{U}(0, 50)$
                & $23.6^{+7.2}_{-6.5}$\\
            $\phi_\mathrm{facu}$ 
                & stellar facula coverage fraction (\%)
                & $\mathcal{U}(0, 50)$
                & $9.6^{+11.7}_{-6.8}$ \\
            \\ \multicolumn{4}{c}{GTC OSIRIS (17 Mar 2013)} \\
            $\Delta T_\mathrm{spot}$
                & $T_\mathrm{spot} - T_\mathrm{phot}$ (K)
                & $\mathcal{U}(-2000, -200)$
                & $-1302^{+380}_{-341}$ \\
            $\Delta T_\mathrm{facu}$ 
                & $T_\mathrm{facu} - T_\mathrm{phot}$ (K)
                & $\mathcal{U}(0, 1000)$
                & $128^{+167}_{-87}$ \\
            $\phi_\mathrm{spot}$ 
                & stellar spot coverage fraction (\%)
                & $\mathcal{U}(0, 50)$
                & $8.8^{+3.1}_{-1.7}$ \\
            $\phi_\mathrm{facu}$ 
                & stellar facula coverage fraction (\%)
                & $\mathcal{U}(0, 50)$
                & $8.2^{+11.3}_{-5.7}$\\
            \\ \multicolumn{4}{c}{LBT MODS (25 Mar 2017)}\\
            $\Delta T_\mathrm{spot}$ 
                & $T_\mathrm{spot} - T_\mathrm{phot}$ (K)
                & $\mathcal{U}(-2000, -200)$
                & $-1679^{+328}_{-217}$\\
            $\Delta T_\mathrm{facu}$
                & $T_\mathrm{facu} - T_\mathrm{phot}$ (K)
                & $\mathcal{U}(0, 1000)$
                & $96^{+118}_{-62}$\\
            $\phi_\mathrm{spot}$
                & stellar spot coverage fraction (\%)
                & $\mathcal{U}(0, 50)$
                & $9.1^{+1.9}_{-1.4}$\\
            $\phi_\mathrm{facu}$
                & stellar facula coverage fraction (\%)
                & $\mathcal{U}(0, 50)$
                & $10.8^{+13.4}_{-7.3}$\\
            \\ \multicolumn{4}{c}{HST STIS G750L (4 Feb 2013)}\\
            $\Delta T_\mathrm{spot}$ 
                & $T_\mathrm{spot} - T_\mathrm{phot}$ (K)
                & $\mathcal{U}(-2000, -200)$
                & $-584^{+244}_{-415}$\\
            $\Delta T_\mathrm{facu}$ 
                & $T_\mathrm{facu} - T_\mathrm{phot}$ (K)
                & $\mathcal{U}(0, 1000)$
                & $163^{+207}_{-111}$\\
            $\phi_\mathrm{spot}$ 
                & stellar spot coverage fraction (\%)
                & $\mathcal{U}(0, 50)$
                & $16.5^{+9.6}_{-6.2}$\\
            $\phi_\mathrm{facu}$ 
                & stellar facula coverage fraction (\%)
                & $\mathcal{U}(0, 50)$
                & $11.7^{+15.0}_{-8.2}$\\
            \\ \multicolumn{4}{c}{HST WFC3 G141 (12 Dec 2015 \& 31 Aug 2016)}\\
            $\beta_{\rm WFC3}$ 
                & transit depth rescaling factor
                & $\mathcal{U}(0.5, 1.5)$
                & $1.0354^{+0.0077}_{-0.0073}$ \\
        \midrule
        \end{tabular}
        
        \end{threeparttable}
    \end{table*}
    
\section{Conclusions}
    \label{sect_conclusions}

    We obtained the optical transmission spectra of HAT-P-12b
    using two transits observed by GTC OSIRIS.
    The derived transit parameters of OB12 and OB13 are consistent with 
    each other and also agree with prior literature values.
    However, there are some systematic biases between 
    the two observed transmission spectra.
    We found that these differences can be attributed to 
    the presence of star spots and faculae.
    We applied the stellar contamination correction 
    as part of the atmospheric retrieval algorithm
    and successfully obtained consistent results of the two transmission spectra.
    
    The retrieved one-dimensional atmospheric model reveals 
    an atmospheric temperature lower than the equilibrium temperature.
    It also results in extremely low abundances of alkali species 
    but high abundances of water, as well as carbon- and nitrogen-bearing gases,
    constrained by the equilibrium chemistry network.
    According to model comparison based on Bayesian evidence,
    our optical transmission spectra provide weak evidence 
    for a clear atmosphere against condensate scattering,
    which is discrepant with previous work based on HST optical and NIR observations
    (\citealt{Sing2016}; \citealt{Wong2020}).
    In addition, we performed a joint retrieval combined with 
    the published transmission spectra of HAT-P-12b,
    coupled with stellar contamination corrections for different observations,
    and the results did not vary, except for a higher metallicity.
    We note that our inferences are obtained under the assumption of stellar contamination,
    which is an alternative explanation for the systematic offset 
    of transmission spectra observed at different transit epochs.
    Although the effect of clouds/hazes and that of stellar activity can be quite similar 
    on transit depths in wavelength ranges from {\it R} band to the infrared,
    such degeneracy can be broken by covering broader ranges 
    with wavelength shorter than 500 nm.
    Unfortunately, the data from HST STIS G430L presented in \citep{Wong2020} 
    could not provide effective constraints on our parametric model of stellar activity.
    
    The study on transit spectroscopy of HAT-P-12b helps us to better understand
    the physical properties and chemical compositions of its atmosphere.
    However, with current low-to-medium-resolution instruments,
    it is challenging to acquire repeatable atmospheric features 
    with high signal-to-noise ratios when there is potential stellar contamination.
    Although we have shown we can reconcile discrepant results 
    from different instruments and different epochs of observations
    via a parametric correction of stellar contamination,
    dozens of free parameters are required to account for 
    stellar activity in multiple datasets,
    which might lead to the ``curse of dimensionality''
    when there are insufficient data constraints.
    Therefore, it is desirable to perform transit spectroscopy 
    that covers a much broader wavelength range in just one transit observation
    so as to avoid concatenating multiple transmission spectra 
    suffering from different levels of stellar contamination.
    Furthermore, a self-consistent model of stellar activity 
    based on multi-band long-term flux monitoring 
    \citep{Rosich2020} is also desired,
    which should be able to predict the intensity 
    of spots and faculae using a limited number of parameters.
    
\begin{acknowledgements}
    G.C. acknowledges the support by the B-type Strategic Priority Program 
    of the Chinese Academy of Sciences (Grant No.\,XDB41000000), 
    the National Natural Science Foundation of China (Grant No.\,42075122, 12122308), 
    the Natural Science Foundation of Jiangsu Province (Grant No.\,BK20190110), 
    Youth Innovation Promotion Association CAS (2021315), 
    and the Minor Planet Foundation of the Purple Mountain Observatory. 
    This work is based on observations made with the Gran Telescopio Canarias (GTC), 
    installed at the Spanish Observatorio del Roque de los Muchachos 
    of the Instituto de Astrof\'{i}sica de Canarias, in the island of La Palma.
    We also thank for the valuable comments and suggestions from the anonymous reviewer.
\end{acknowledgements}

\bibliographystyle{aa}
\bibliography{reference}

\begin{thebibliography}{64}
\expandafter\ifx\csname natexlab\endcsname\relax\def\natexlab#1{#1}\fi

\bibitem[{{Alexoudi} {et~al.}(2020){Alexoudi}, {Mallonn}, {Keles},
  {Poppenh{\"a}ger}, {von Essen}, \& {Strassmeier}}]{Alexoudi2020}
{Alexoudi}, X., {Mallonn}, M., {Keles}, E., {et~al.} 2020, \aap, 640, A134

\bibitem[{{Alexoudi} {et~al.}(2018){Alexoudi}, {Mallonn}, {von Essen},
  {Turner}, {Keles}, {Southworth}, {Mancini}, {Ciceri}, {Granzer}, {Denker},
  {Dineva}, \& {Strassmeier}}]{Alexoudi2018}
{Alexoudi}, X., {Mallonn}, M., {von Essen}, C., {et~al.} 2018, \aap, 620, A142

\bibitem[{{Allard} {et~al.}(2012){Allard}, {Homeier}, \&
  {Freytag}}]{Allard2012}
{Allard}, F., {Homeier}, D., \& {Freytag}, B. 2012, Philosophical Transactions
  of the Royal Society of London Series A, 370, 2765

\bibitem[{{Barstow} {et~al.}(2017){Barstow}, {Aigrain}, {Irwin}, \&
  {Sing}}]{Barstow2017}
{Barstow}, J.~K., {Aigrain}, S., {Irwin}, P.~G.~J., \& {Sing}, D.~K. 2017,
  \apj, 834, 50

\bibitem[{{Benneke} {et~al.}(2019){Benneke}, {Knutson}, {Lothringer},
  {Crossfield}, {Moses}, {Morley}, {Kreidberg}, {Fulton}, {Dragomir}, {Howard},
  {Wong}, {D{\'e}sert}, {McCullough}, {Kempton}, {Fortney}, {Gilliland},
  {Deming}, \& {Kammer}}]{Benneke2019}
{Benneke}, B., {Knutson}, H.~A., {Lothringer}, J., {et~al.} 2019, Nature
  Astronomy, 3, 813

\bibitem[{{Berdyugina}(2005)}]{Berdyugina2005}
{Berdyugina}, S.~V. 2005, Living Reviews in Solar Physics, 2, 8

\bibitem[{{Brown}(2001)}]{Brown2001}
{Brown}, T.~M. 2001, \apj, 553, 1006

\bibitem[{{Buchner} {et~al.}(2014){Buchner}, {Georgakakis}, {Nandra}, {Hsu},
  {Rangel}, {Brightman}, {Merloni}, {Salvato}, {Donley}, \&
  {Kocevski}}]{code-PyMultiNest}
{Buchner}, J., {Georgakakis}, A., {Nandra}, K., {et~al.} 2014, \aap, 564, A125

\bibitem[{{Cepa} {et~al.}(2000){Cepa}, {Aguiar}, {Escalera},
  {Gonzalez-Serrano}, {Joven-Alvarez}, {Peraza}, {Rasilla}, {Rodriguez-Ramos},
  {Gonzalez}, {Cobos Duenas}, {Sanchez}, {Tejada}, {Bland-Hawthorn},
  {Militello}, \& {Rosa}}]{OSIRIS2000}
{Cepa}, J., {Aguiar}, M., {Escalera}, V.~G., {et~al.} 2000, in Society of
  Photo-Optical Instrumentation Engineers (SPIE) Conference Series, Vol. 4008,
  Optical and IR Telescope Instrumentation and Detectors, ed. M.~{Iye} \& A.~F.
  {Moorwood}, 623--631

\bibitem[{{Chapman} {et~al.}(1997){Chapman}, {Cookson}, \&
  {Dobias}}]{Chapman1997}
{Chapman}, G.~A., {Cookson}, A.~M., \& {Dobias}, J.~J. 1997, \apj, 482, 541

\bibitem[{{Chen} {et~al.}(2017){Chen}, {Guenther}, {Pall{\'e}}, {Nortmann},
  {Nowak}, {Kunz}, {Parviainen}, \& {Murgas}}]{Chen2017a}
{Chen}, G., {Guenther}, E.~W., {Pall{\'e}}, E., {et~al.} 2017, \aap, 600, A138

\bibitem[{{Deibert} {et~al.}(2019){Deibert}, {de Mooij}, {Jayawardhana},
  {Fortney}, {Brogi}, {Rustamkulov}, \& {Tamura}}]{Deibert2019}
{Deibert}, E.~K., {de Mooij}, E. J.~W., {Jayawardhana}, R., {et~al.} 2019, \aj,
  157, 58

\bibitem[{{Deming} {et~al.}(2013){Deming}, {Wilkins}, {McCullough}, {Burrows},
  {Fortney}, {Agol}, {Dobbs-Dixon}, {Madhusudhan}, {Crouzet}, {Desert},
  {Gilliland}, {Haynes}, {Knutson}, {Line}, {Magic}, {Mandell}, {Ranjan},
  {Charbonneau}, {Clampin}, {Seager}, \& {Showman}}]{Deming2013}
{Deming}, D., {Wilkins}, A., {McCullough}, P., {et~al.} 2013, \apj, 774, 95

\bibitem[{{Espinoza} \& {Jord{\'a}n}(2015)}]{code-limb-darkening}
{Espinoza}, N. \& {Jord{\'a}n}, A. 2015, \mnras, 450, 1879

\bibitem[{{Evans} {et~al.}(2016){Evans}, {Sing}, {Wakeford}, {Nikolov},
  {Ballester}, {Drummond}, {Kataria}, {Gibson}, {Amundsen}, \&
  {Spake}}]{Evans2016}
{Evans}, T.~M., {Sing}, D.~K., {Wakeford}, H.~R., {et~al.} 2016, \apjl, 822, L4

\bibitem[{{Feroz} \& {Hobson}(2008)}]{Feroz2008}
{Feroz}, F. \& {Hobson}, M.~P. 2008, \mnras, 384, 449

\bibitem[{{Feroz} {et~al.}(2009){Feroz}, {Hobson}, \& {Bridges}}]{Feroz2009}
{Feroz}, F., {Hobson}, M.~P., \& {Bridges}, M. 2009, \mnras, 398, 1601

\bibitem[{{Fisher} \& {Heng}(2018)}]{Fisher2018}
{Fisher}, C. \& {Heng}, K. 2018, \mnras, 481, 4698

\bibitem[{{Foreman-Mackey} {et~al.}(2017){Foreman-Mackey}, {Agol}, {Angus}, \&
  {Ambikasaran}}]{code-celerite}
{Foreman-Mackey}, D., {Agol}, E., {Angus}, R., \& {Ambikasaran}, S. 2017, ArXiv

\bibitem[{{Gibson} {et~al.}(2012){Gibson}, {Aigrain}, {Roberts}, {Evans},
  {Osborne}, \& {Pont}}]{Gibson2012}
{Gibson}, N.~P., {Aigrain}, S., {Roberts}, S., {et~al.} 2012, \mnras, 419, 2683

\bibitem[{{Gibson} {et~al.}(2017){Gibson}, {Nikolov}, {Sing}, {Barstow},
  {Evans}, {Kataria}, \& {Wilson}}]{Gibson2017}
{Gibson}, N.~P., {Nikolov}, N., {Sing}, D.~K., {et~al.} 2017, \mnras, 467, 4591

\bibitem[{{Gordon} {et~al.}(2017){Gordon}, {Rothman}, {Hill}, {Kochanov},
  {Tan}, {Bernath}, {Birk}, {Boudon}, {Campargue}, {Chance}, {Drouin}, {Flaud},
  {Gamache}, {Hodges}, {Jacquemart}, {Perevalov}, {Perrin}, {Shine}, {Smith},
  {Tennyson}, {Toon}, {Tran}, {Tyuterev}, {Barbe}, {Cs{\'a}sz{\'a}r}, {Devi},
  {Furtenbacher}, {Harrison}, {Hartmann}, {Jolly}, {Johnson}, {Karman},
  {Kleiner}, {Kyuberis}, {Loos}, {Lyulin}, {Massie}, {Mikhailenko},
  {Moazzen-Ahmadi}, {M{\"u}ller}, {Naumenko}, {Nikitin}, {Polyansky}, {Rey},
  {Rotger}, {Sharpe}, {Sung}, {Starikova}, {Tashkun}, {Auwera}, {Wagner},
  {Wilzewski}, {Wcis{\l}o}, {Yu}, \& {Zak}}]{HITRAN2016}
{Gordon}, I.~E., {Rothman}, L.~S., {Hill}, C., {et~al.} 2017, \jqsrt, 203, 3

\bibitem[{{Greene} {et~al.}(2016){Greene}, {Line}, {Montero}, {Fortney},
  {Lustig-Yaeger}, \& {Luther}}]{Greene2016}
{Greene}, T.~P., {Line}, M.~R., {Montero}, C., {et~al.} 2016, \apj, 817, 17

\bibitem[{{Hartman} {et~al.}(2009){Hartman}, {Bakos}, {Torres}, {Kov{\'a}cs},
  {Noyes}, {P{\'a}l}, {Latham}, {Sip{\H{o}}cz}, {Fischer}, {Johnson}, {Marcy},
  {Butler}, {Howard}, {Esquerdo}, {Sasselov}, {Kov{\'a}cs}, {Stefanik},
  {Fernandez}, {L{\'a}z{\'a}r}, {Papp}, \& {S{\'a}ri}}]{confirm-HATP12b}
{Hartman}, J.~D., {Bakos}, G.~{\'A}., {Torres}, G., {et~al.} 2009, \apj, 706,
  785

\bibitem[{{Herrero} {et~al.}(2016){Herrero}, {Ribas}, {Jordi}, {Morales},
  {Perger}, \& {Rosich}}]{Herrero2016}
{Herrero}, E., {Ribas}, I., {Jordi}, C., {et~al.} 2016, \aap, 586, A131

\bibitem[{{Johnson} {et~al.}(2021){Johnson}, {Norris}, {Unruh}, {Solanki},
  {Krivova}, {Witzke}, \& {Shapiro}}]{Johnson2021}
{Johnson}, L.~J., {Norris}, C.~M., {Unruh}, Y.~C., {et~al.} 2021, \mnras
  [\eprint[arXiv]{2104.11544}]

\bibitem[{{Kass} \& {Raftery}(1995)}]{Kass1995}
{Kass}, R.~E. \& {Raftery}, A.~E. 1995, Journal of the American Statistical
  Association, 90, 773

\bibitem[{{Kitzmann} \& {Heng}(2018)}]{Kitzmann2018}
{Kitzmann}, D. \& {Heng}, K. 2018, \mnras, 475, 94

\bibitem[{{Kreidberg}(2015)}]{code-batman}
{Kreidberg}, L. 2015, \pasp, 127, 1161

\bibitem[{{Kreidberg} {et~al.}(2014){Kreidberg}, {Bean}, {D{\'e}sert},
  {Benneke}, {Deming}, {Stevenson}, {Seager}, {Berta-Thompson}, {Seifahrt}, \&
  {Homeier}}]{Kreidberg2014Nature}
{Kreidberg}, L., {Bean}, J.~L., {D{\'e}sert}, J.-M., {et~al.} 2014, \nat, 505,
  69

\bibitem[{{Line} {et~al.}(2013){Line}, {Knutson}, {Deming}, {Wilkins}, \&
  {Desert}}]{Line2013}
{Line}, M.~R., {Knutson}, H., {Deming}, D., {Wilkins}, A., \& {Desert}, J.-M.
  2013, \apj, 778, 183

\bibitem[{{MacDonald} {et~al.}(2020){MacDonald}, {Goyal}, \&
  {Lewis}}]{MacDonald2020}
{MacDonald}, R.~J., {Goyal}, J.~M., \& {Lewis}, N.~K. 2020, \apjl, 893, L43

\bibitem[{{Madhusudhan} \& {Seager}(2009)}]{Madhusudhan2009}
{Madhusudhan}, N. \& {Seager}, S. 2009, \apj, 707, 24

\bibitem[{{Mallonn} {et~al.}(2015){Mallonn}, {Nascimbeni}, {Weingrill}, {von
  Essen}, {Strassmeier}, {Piotto}, {Pagano}, {Scandariato}, {Csizmadia},
  {Herrero}, {Sada}, {Dhillon}, {Marsh}, {K{\"u}nstler}, {Bernt}, \&
  {Granzer}}]{Mallonn2015}
{Mallonn}, M., {Nascimbeni}, V., {Weingrill}, J., {et~al.} 2015, \aap, 583,
  A138

\bibitem[{{Mancini} {et~al.}(2018){Mancini}, {Esposito}, {Covino},
  {Southworth}, {Biazzo}, {Bruni}, {Ciceri}, {Evans}, {Lanza}, {Poretti},
  {Sarkis}, {Smith}, {Brogi}, {Affer}, {Benatti}, {Bignamini}, {Boccato},
  {Bonomo}, {Borsa}, {Carleo}, {Claudi}, {Cosentino}, {Damasso}, {Desidera},
  {Giacobbe}, {Gonz{\'a}lez-{\'A}lvarez}, {Gratton}, {Harutyunyan}, {Leto},
  {Maggio}, {Malavolta}, {Maldonado}, {Martinez-Fiorenzano}, {Masiero},
  {Micela}, {Molinari}, {Nascimbeni}, {Pagano}, {Pedani}, {Piotto}, {Rainer},
  {Scandariato}, {Smareglia}, {Sozzetti}, {Andreuzzi}, \&
  {Henning}}]{Mancini2018}
{Mancini}, L., {Esposito}, M., {Covino}, E., {et~al.} 2018, \aap, 613, A41

\bibitem[{{Mandel} \& {Agol}(2002)}]{Mandel2002}
{Mandel}, K. \& {Agol}, E. 2002, \apjl, 580, L171

\bibitem[{{McCullough} {et~al.}(2014){McCullough}, {Crouzet}, {Deming}, \&
  {Madhusudhan}}]{McCullough2014}
{McCullough}, P.~R., {Crouzet}, N., {Deming}, D., \& {Madhusudhan}, N. 2014,
  \apj, 791, 55

\bibitem[{{Nortmann} {et~al.}(2016){Nortmann}, {Pall{\'e}}, {Murgas},
  {Dreizler}, {Iro}, \& {Cabrera-Lavers}}]{GTC-HATP32b}
{Nortmann}, L., {Pall{\'e}}, E., {Murgas}, F., {et~al.} 2016, \aap, 594, A65

\bibitem[{{Pinhas} {et~al.}(2019){Pinhas}, {Madhusudhan}, {Gandhi}, \&
  {MacDonald}}]{Pinhas2019}
{Pinhas}, A., {Madhusudhan}, N., {Gandhi}, S., \& {MacDonald}, R. 2019, \mnras,
  482, 1485

\bibitem[{{Pinhas} {et~al.}(2018){Pinhas}, {Rackham}, {Madhusudhan}, \&
  {Apai}}]{Pinhas2018}
{Pinhas}, A., {Rackham}, B.~V., {Madhusudhan}, N., \& {Apai}, D. 2018, \mnras,
  480, 5314

\bibitem[{{Pont} {et~al.}(2013){Pont}, {Sing}, {Gibson}, {Aigrain}, {Henry}, \&
  {Husnoo}}]{Pont2013}
{Pont}, F., {Sing}, D.~K., {Gibson}, N.~P., {et~al.} 2013, \mnras, 432, 2917

\bibitem[{{Rackham} {et~al.}(2018){Rackham}, {Apai}, \&
  {Giampapa}}]{Rackham2018}
{Rackham}, B.~V., {Apai}, D., \& {Giampapa}, M.~S. 2018, \apj, 853, 122

\bibitem[{{Rackham} {et~al.}(2019){Rackham}, {Apai}, \&
  {Giampapa}}]{Rackham2019}
{Rackham}, B.~V., {Apai}, D., \& {Giampapa}, M.~S. 2019, \aj, 157, 96

\bibitem[{{Rasmussen} \& {Williams}(2006)}]{Rasmussen2006}
{Rasmussen}, C.~E. \& {Williams}, C. K.~I. 2006, {Gaussian Processes for
  Machine Learning} ({The MIT Press})

\bibitem[{{Rey} {et~al.}(2017){Rey}, {Nikitin}, \& {Tyuterev}}]{Rey2017}
{Rey}, M., {Nikitin}, A.~V., \& {Tyuterev}, V.~G. 2017, \apj, 847, 105

\bibitem[{{Richard} {et~al.}(2012){Richard}, {Gordon}, {Rothman}, {Abel},
  {Frommhold}, {Gustafsson}, {Hartmann}, {Hermans}, {Lafferty}, {Orton},
  {Smith}, \& {Tran}}]{HITRAN2012}
{Richard}, C., {Gordon}, I.~E., {Rothman}, L.~S., {et~al.} 2012, \jqsrt, 113,
  1276

\bibitem[{{Rosich} {et~al.}(2020){Rosich}, {Herrero}, {Mallonn}, {Ribas},
  {Morales}, {Perger}, {Anglada-Escud{\'e}}, \& {Granzer}}]{Rosich2020}
{Rosich}, A., {Herrero}, E., {Mallonn}, M., {et~al.} 2020, \aap, 641, A82

\bibitem[{{Sansonetti} \& {Martin}(2005)}]{NIST}
{Sansonetti}, J.~E. \& {Martin}, W.~C. 2005, Journal of Physical and Chemical
  Reference Data, 34, 1559

\bibitem[{{Schlawin} {et~al.}(2018){Schlawin}, {Greene}, {Line}, {Fortney}, \&
  {Rieke}}]{Schlawin2018}
{Schlawin}, E., {Greene}, T.~P., {Line}, M., {Fortney}, J.~J., \& {Rieke}, M.
  2018, \aj, 156, 40

\bibitem[{{Seager} \& {Sasselov}(2000)}]{Seager2000}
{Seager}, S. \& {Sasselov}, D.~D. 2000, \apj, 537, 916

\bibitem[{{Sedaghati} {et~al.}(2017){Sedaghati}, {Boffin}, {MacDonald},
  {Gandhi}, {Madhusudhan}, {Gibson}, {Oshagh}, {Claret}, \&
  {Rauer}}]{Sedaghati2017}
{Sedaghati}, E., {Boffin}, H. M.~J., {MacDonald}, R.~J., {et~al.} 2017, \nat,
  549, 238

\bibitem[{{Sing} {et~al.}(2016){Sing}, {Fortney}, {Nikolov}, {Wakeford},
  {Kataria}, {Evans}, {Aigrain}, {Ballester}, {Burrows}, {Deming},
  {D{\'e}sert}, {Gibson}, {Henry}, {Huitson}, {Knutson}, {Lecavelier Des
  Etangs}, {Pont}, {Showman}, {Vidal-Madjar}, {Williamson}, \&
  {Wilson}}]{Sing2016}
{Sing}, D.~K., {Fortney}, J.~J., {Nikolov}, N., {et~al.} 2016, \nat, 529, 59

\bibitem[{{Skilling}(2004)}]{Skilling2004}
{Skilling}, J. 2004, in American Institute of Physics Conference Series, Vol.
  735, Bayesian Inference and Maximum Entropy Methods in Science and
  Engineering: 24th International Workshop on Bayesian Inference and Maximum
  Entropy Methods in Science and Engineering, ed. R.~{Fischer}, R.~{Preuss}, \&
  U.~V. {Toussaint}, 395--405

\bibitem[{{Tashkun} \& {Perevalov}(2011)}]{CDSD4000}
{Tashkun}, S.~A. \& {Perevalov}, V.~I. 2011, \jqsrt, 112, 1403

\bibitem[{{Tennyson} \& {Yurchenko}(2018)}]{ExoMol}
{Tennyson}, J. \& {Yurchenko}, S. 2018, Atoms, 6, 26

\bibitem[{{Thorngren} {et~al.}(2016){Thorngren}, {Fortney}, {Murray-Clay}, \&
  {Lopez}}]{Thorngren2016}
{Thorngren}, D.~P., {Fortney}, J.~J., {Murray-Clay}, R.~A., \& {Lopez}, E.~D.
  2016, \apj, 831, 64

\bibitem[{{Tody}(1993)}]{code-IRAF}
{Tody}, D. 1993, in Astronomical Society of the Pacific Conference Series,
  Vol.~52, Astronomical Data Analysis Software and Systems II, ed. R.~J.
  {Hanisch}, R.~J.~V. {Brissenden}, \& J.~{Barnes}, 173

\bibitem[{{Tsiaras} {et~al.}(2018){Tsiaras}, {Waldmann}, {Zingales},
  {Rocchetto}, {Morello}, {Damiano}, {Karpouzas}, {Tinetti}, {McKemmish},
  {Tennyson}, \& {Yurchenko}}]{Tsiaras2018}
{Tsiaras}, A., {Waldmann}, I.~P., {Zingales}, T., {et~al.} 2018, \aj, 155, 156

\bibitem[{{Woitke} {et~al.}(2018){Woitke}, {Helling}, {Hunter}, {Millard},
  {Turner}, {Worters}, {Blecic}, \& {Stock}}]{code-GGChem}
{Woitke}, P., {Helling}, C., {Hunter}, G.~H., {et~al.} 2018, \aap, 614, A1

\bibitem[{{Wong} {et~al.}(2020){Wong}, {Benneke}, {Gao}, {Knutson}, {Chachan},
  {Henry}, {Deming}, {Kataria}, {Lee}, {Nikolov}, {Sing}, {Ballester},
  {Baskin}, {Wakeford}, \& {Williamson}}]{Wong2020}
{Wong}, I., {Benneke}, B., {Gao}, P., {et~al.} 2020, \aj, 159, 234

\bibitem[{{Yan} {et~al.}(2020){Yan}, {Espinoza}, {Molaverdikhani}, {Henning},
  {Mancini}, {Mallonn}, {Rackham}, {Apai}, {Jord{\'a}n}, {Molli{\`e}re},
  {Chen}, {Carone}, \& {Reiners}}]{Yan2020}
{Yan}, F., {Espinoza}, N., {Molaverdikhani}, K., {et~al.} 2020, \aap, 642, A98

\bibitem[{{Zacharias} {et~al.}(2015){Zacharias}, {Finch}, {Subasavage},
  {Bredthauer}, {Crockett}, {Divittorio}, {Ferguson}, {Harris}, {Harris},
  {Henden}, {Kilian}, {Munn}, {Rafferty}, {Rhodes}, {Schultheiss}, {Tilleman},
  \& {Wieder}}]{URAT1}
{Zacharias}, N., {Finch}, C., {Subasavage}, J., {et~al.} 2015, \aj, 150, 101

\bibitem[{{Zhang} \& {Chachan}(2019)}]{code-PLATON}
{Zhang}, M. \& {Chachan}, Y. 2019, {PLATON: PLanetary Atmospheric Transmission
  for Observer Noobs}

\bibitem[{{Zhang} {et~al.}(2020){Zhang}, {Chachan}, {Kempton}, {Knutson}, \&
  {Chang}}]{code-PLATON2}
{Zhang}, M., {Chachan}, Y., {Kempton}, E. M.~R., {Knutson}, H.~A., \& {Chang},
  W.~H. 2020, \apj, 899, 27

\end{thebibliography}

\begin{appendix}
    
\section{Appendix}
    
    \begin{figure*}[bp]
        \centering
        \includegraphics[width=\linewidth]{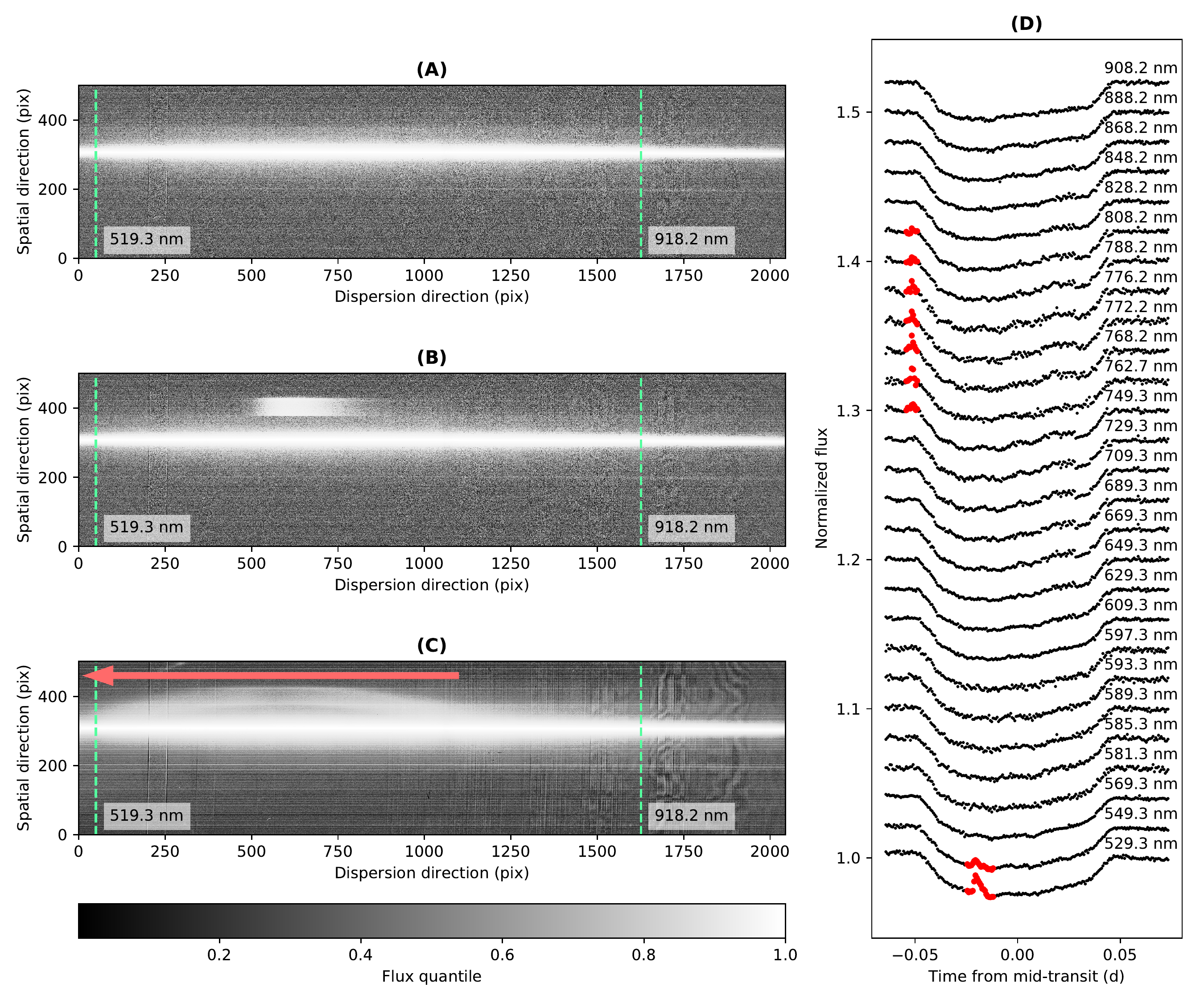}
        \caption{A moving ghost contaminating the spectral images of OB12. 
        Panel A: a clear frame near the mid-transit. 
        Panel B: a contaminated frame near the ingress of the transit.
        Panel C: stacking all contaminated frames showing the trace of ghost.
        The green dashed lines indicate the waveband for white-light curves.
        The red arrow indicates the moving direction of the ghost from the redder side to the bluer side.
        Panel D: raw spectroscopic light curves used in Sect. \ref{sect_spectroscopic_light_curves}.
        The red dots correspond to the contaminated frames,
        which were removed in the spectroscopic light curve fitting.
        We note that the horizontal moire fringes shown in Panel A, B and C
        were caused by the 500-kHz fast readout mode adopted in OB12
        and were visually magnified due to a small figure size,
        which barely affected stellar fluxes.}
        \label{fig_ghost}
    \end{figure*}
    
    \begin{figure*}
        \centering
        \includegraphics[width=\linewidth]{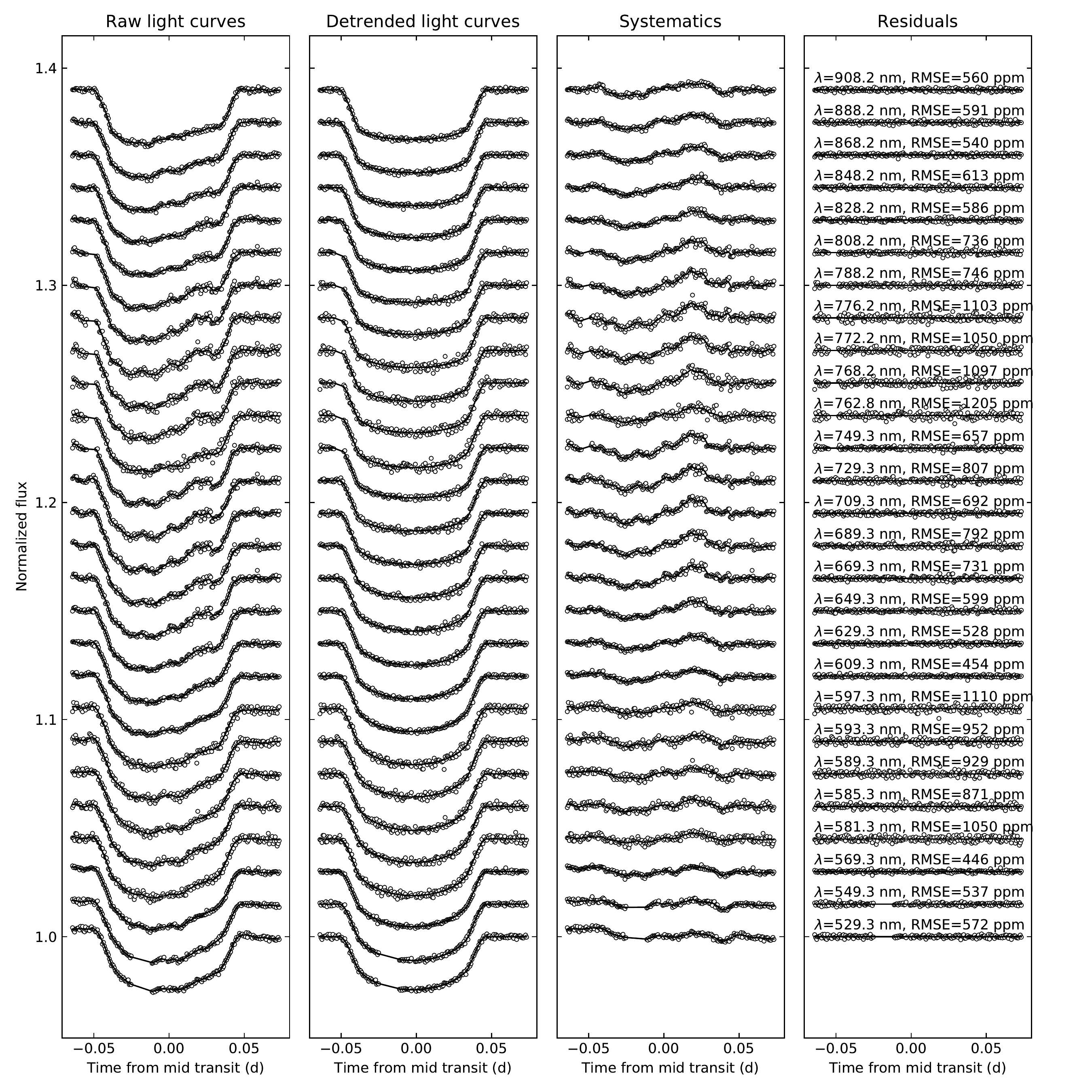}
        \caption{Spectroscopic light curve fitting of OB12. 
        From left to right, 
        the first panel: raw light curves (circles) 
        and the best-fit curves (solid lines);
        the second panel: detrended light curves (circles) 
        and the best-fit transit models (solid lines);
        the third panel: systematics (circles) 
        and their GP models (solid lines);
        the fourth panel: residuals.
        The points affected by the moving ghost have been removed.
        The curves were arbitrarily shifted for clarity.
        }
        \label{fig_specLCs_OB12}
    \end{figure*}    

    \begin{figure*}
        \centering
        \includegraphics[width=\linewidth]{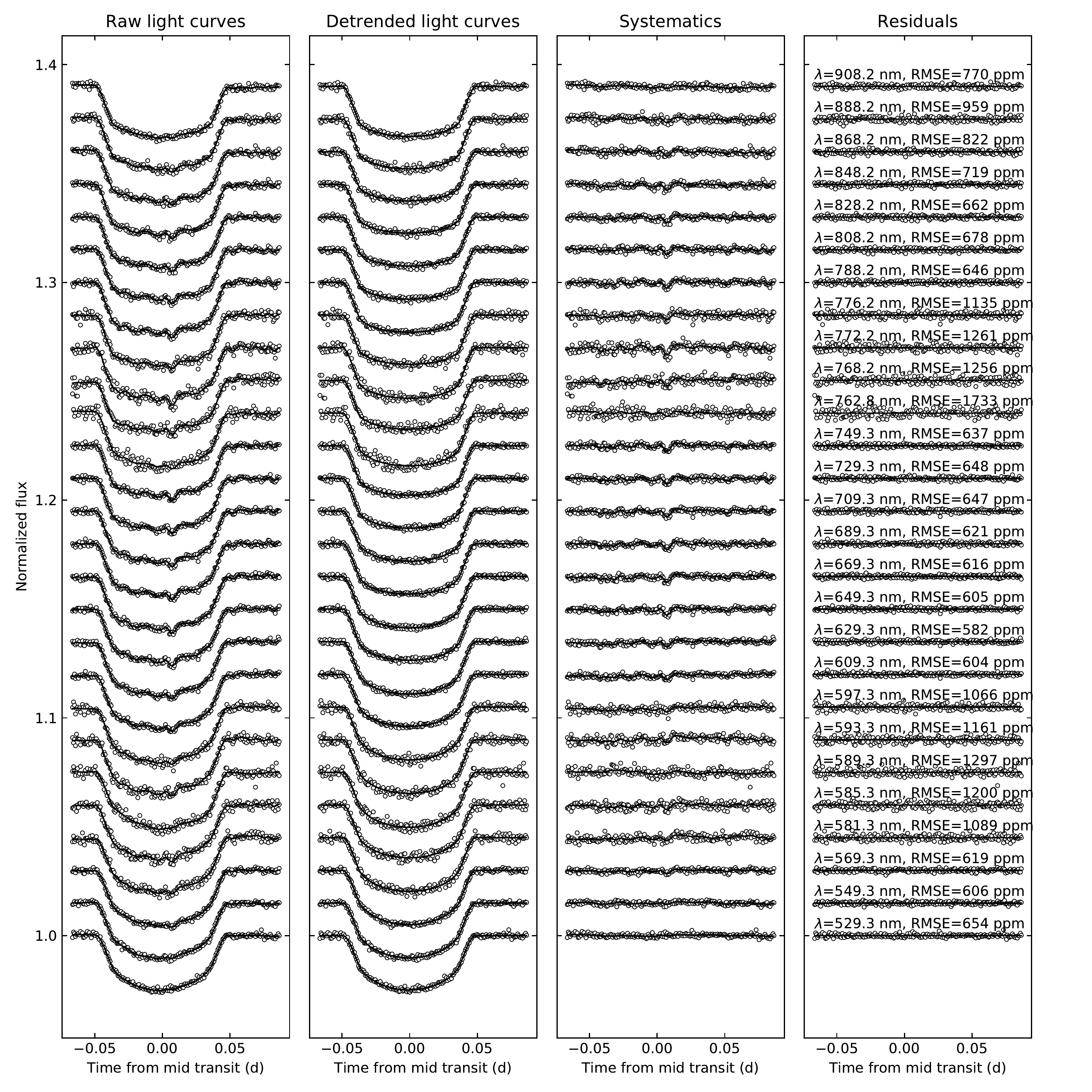}
        \caption{Same as Fig. \ref{fig_specLCs_OB12}, 
        but for light curves from the OB13 dataset.}
        \label{fig_specLCs_OB13}
    \end{figure*}   
    
    \begin{figure*}
        \centering
        \includegraphics[width=\linewidth]{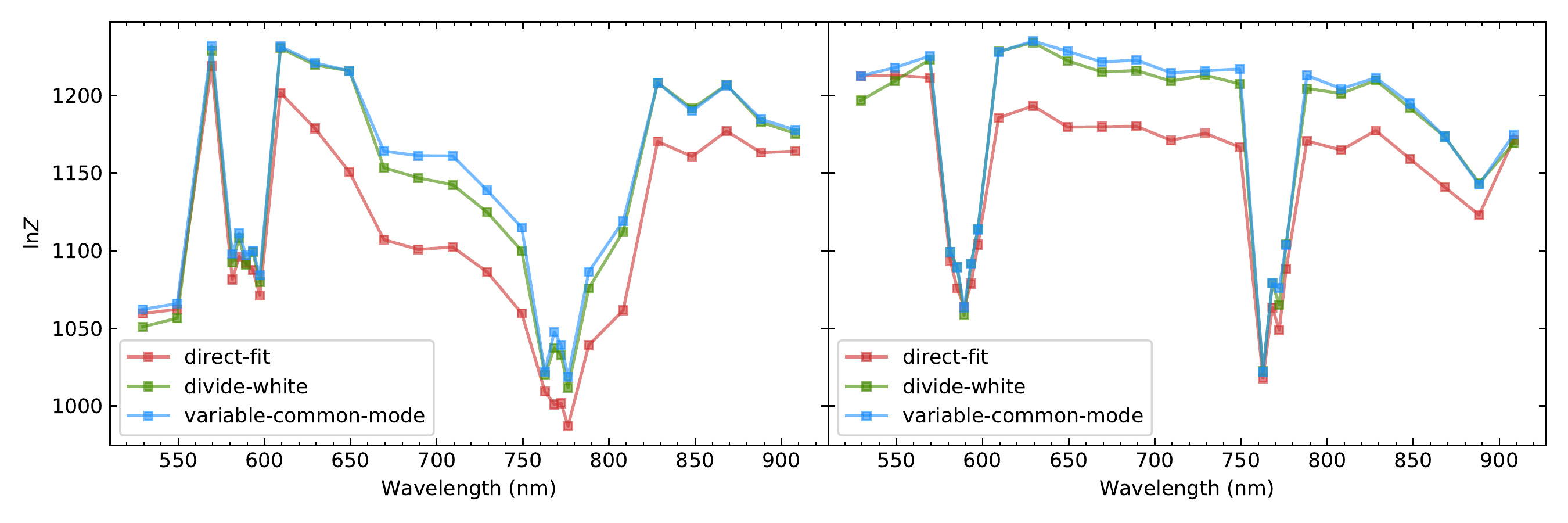}
        \caption{Bayesian evidence for different methods of spectroscopic light curve fitting
            for OB12 (left) and OB13 (right).
            The \texttt{direct-fit} method serves as a control group
            where the common modes are not removed in spectroscopic light curve fitting.}
        \label{fig_lc_evidence}
    \end{figure*}  

    \begin{figure*}
        \centering
        \includegraphics[width=0.8\linewidth]{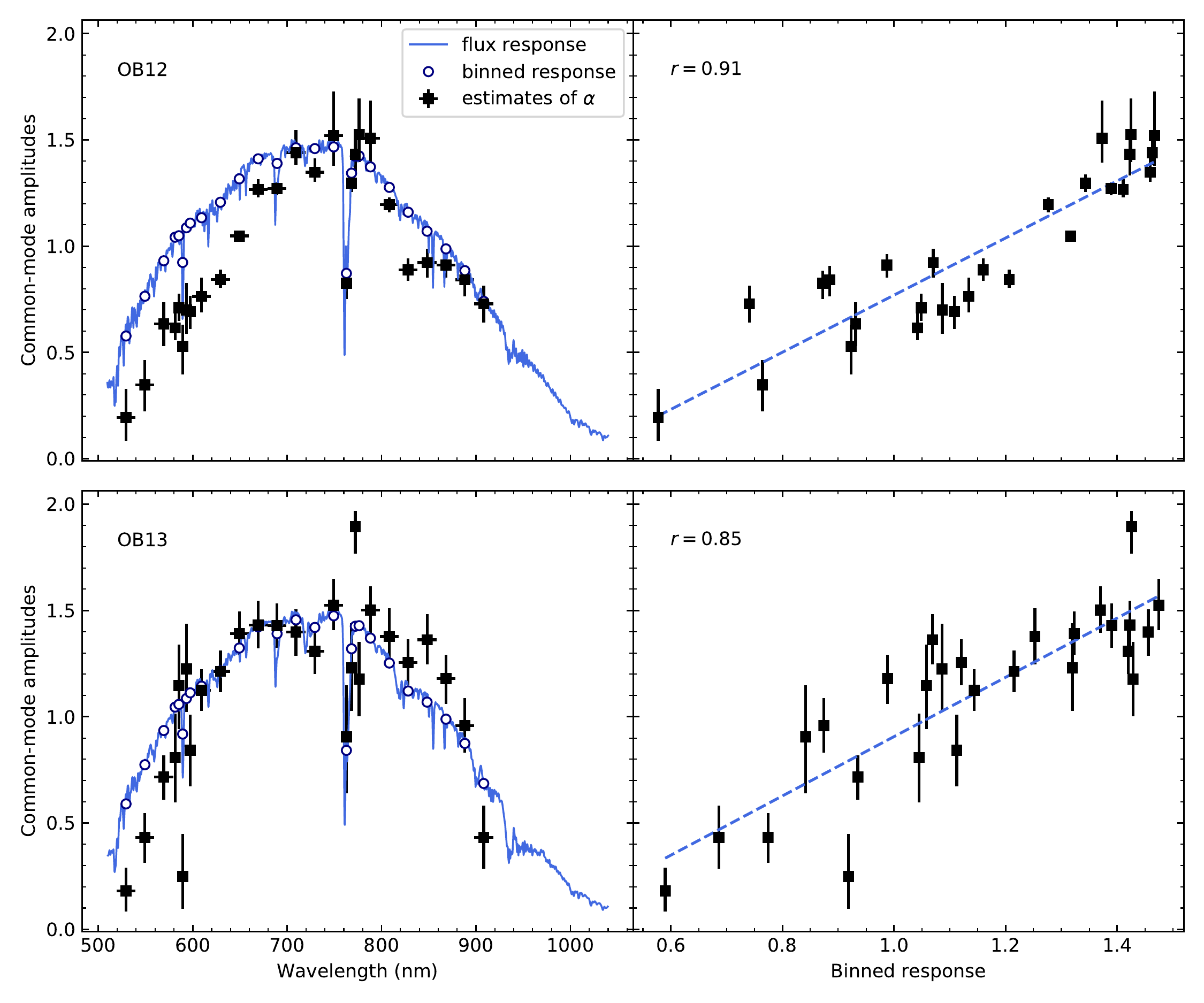}
        \caption{Flux-correlated amplitudes of common modes. 
        The upper and lower panels correspond to the results of OB12 and OB13, respectively.
        Left panels: comparison between the common-mode amplitudes and flux response curves.
        Right panels: correlation between the common-mode amplitudes and binned flux responses.
        The blue dashed lines are linear regression models.
        The Pearson correlation coefficient is denoted as $r$.
        The flux response curves were arbitrarily rescaled for clarity
        but would not affect the correlation.}
        \label{fig_variable_amplitudes}
    \end{figure*}  
    
    \begin{figure*}
        \centering
        \includegraphics[width=\linewidth]{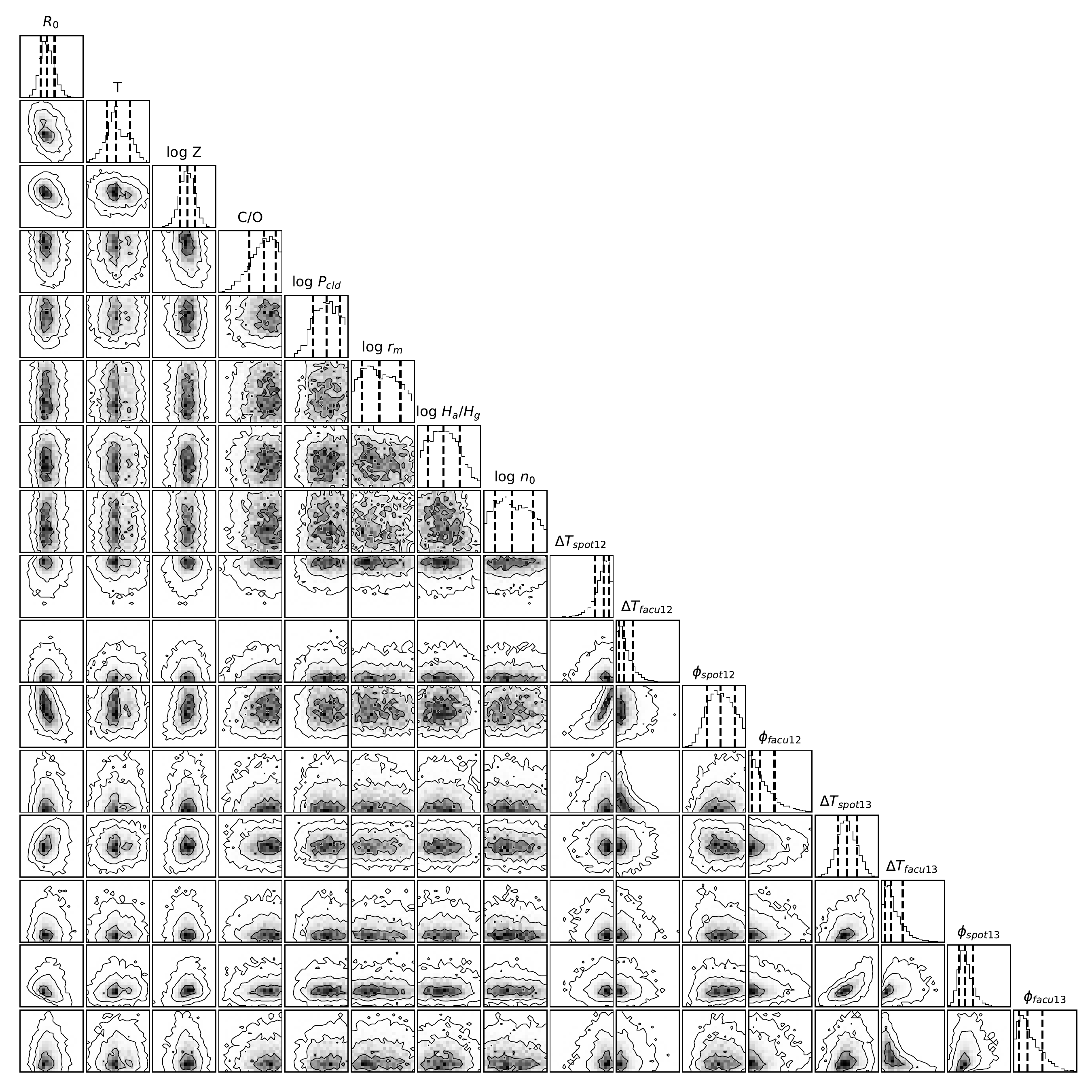}
        \caption{
        Posterior joint distributions of parameters for atmospheric retrieval.
        The contours indicate 39.3\%, 86.5\%, 98.9\% (1- to 3-$\sigma$) credible intervals.
        The diagonal panels show the marginal distributions of corresponding parameters,
        in which the vertical dashed lines indicate the medians and 68.2\% credible intervals.
        The specific values of posteriors are listed in Table \ref{table_retrieval_estimates}.
        }
        \label{fig_retrieval_corners}
    \end{figure*}

\end{appendix}

\end{document}